\documentclass{aa}  

\usepackage{graphicx}
\usepackage{txfonts}
\usepackage{bbm}
\usepackage{url}
\usepackage{natbib}
\usepackage[utf8]{inputenc}
\usepackage[breaklinks=true,draft]{hyperref}
\hypersetup{colorlinks=true,urlcolor=blue,citecolor=blue,pdfborder= 0 0 0}
\usepackage{colortbl, color}

\DeclareUnicodeCharacter{00B0}{\degree} 
\bibpunct{(}{)}{;}{a}{}{,}    

\usepackage{xcolor}

\newcommand{\PP}{{\mathbb P}}

\usepackage{amstext}

\begin{document} 

        \title{Photometric redshift galaxies as tracers of the filamentary network}
        \titlerunning{Photometric redshift galaxies for filament detection}
   \subtitle{}

   \author{M. Kruuse
          \inst{1}
          \and
          E. Tempel\inst{1, 2}
                    \and
          R. Kipper \inst{1}
          \and
          R. S. Stoica \inst{3}
          }

   \institute{
              Tartu Observatory, University of Tartu, Observatooriumi 1, 61602 Tõravere, Estonia
         \and
             Leibniz-Institut für Astrophysik Potsdam (AIP), An der Sternwarte 16, 14482 Potsdam, Germany
                \and
                        Université de Lorraine, CNRS, IECL, F-54000 Nancy, France         
             }

  \abstract
  {Galaxy filaments are the dominant feature in the overall structure of the cosmic web. The study of the filamentary web is an important aspect in understanding galaxy evolution and the evolution of matter in the Universe. A map of the filamentary structure is an adequate probe of the web.}
   {We propose that photometric redshift galaxies are significantly positively associated with the filamentary structure detected from the spatial distribution of spectroscopic redshift galaxies. The long-term aim is to use the photometric galaxies in addition to spectroscopic galaxies to create a more detailed and far-reaching map of the filamentary structure.} 
   {The catalogues of spectroscopic and photometric galaxies are seen as point-process realisations in a sphere, and the catalogue of filamentary spines is proposed to be a realisation of a random set in a sphere. The positive association between these sets was studied using a bivariate $J-$function, which is a summary statistics studying clustering.  A quotient $D$ was built to estimate the distance distribution of the filamentary spine to galaxies in comparison to the distance distribution of the filamentary spine to random points in $3-$dimensional Euclidean space. This measure also gives a physical distance scale to the distances between filamentary spines and the studied sets of galaxies. In the $3-$dimensional case, galaxies and random points are described as line of sights following through their positions on the sphere.} 
   {The bivariate $J-$function shows a statistically significant clustering effect in between filamentary spines and photometric redshift galaxies. The quotient $D$ confirms the previous result that smaller distances exist with higher probability between the photometric galaxies and filaments. The trend of  smaller distances between the objects grows stronger at higher redshift. Additionally, the quotient $D$ for photometric galaxies gives a rough estimate for the filamentary spine width of about $1$~Mpc.}
   {Photometric redshift galaxies are positively associated with filamentary spines detected from the spatial distribution of spectroscopic galaxies. In addition to the spatial distribution of spectroscopic galaxies, the information embedded in the photometric galaxies could contribute greatly to the detection of cosmic web structures.} 

   \keywords{Methods: statistical -- galaxies: statistics -- large-scale structure of Universe}

   \maketitle
%

\section{Introduction}
Galaxies and matter are not uniformly distributed  in the Universe~\citep{martinez2001statistics}. The overall structure of the cosmic web is complex and consists of multiple elements that form the overall picture~\citep{2017prcobookBJ}. 

Large-scale galaxy surveys reveal that galaxies form structures: long string-like filamentary patterns, galaxy clusters, sheets, and superclusters~\citep{joeveer1978spatial, 1996Natur.380..603B}. 
Large-scale galaxy distribution maps are dominated by bridging structures that are called filaments. These connect clusters and border near-empty voids~\citep{2005MNRAS.357L..45P}. The evolution of matter in the Universe depends on the environment it inhabits. \citet{odekon2017effect} studied the cold gas reservoirs in galaxies with similar properties, but varying large-scale structure elements that they belonged to. The authors detected that the cold gas deficiency of certain types of galaxies decreases with distance from the filamentary spine, and most gas-rich galaxies of certain parameters are located in tendrils within voids (see also~\citealt{Kuutma2017}).
Studying the overall structure of the cosmic web is necessary to understand the complicated mechanics inside the web elements and the evolution of galaxies therein.

The formation of galaxies and the further evolution of matter in the Universe has been driven by cosmic fluctuations and gravitational collapse~\citep{Sunyaev1969}. In the Zeldovich pancake picture of the cosmic evolution, matter collapses intrinsically along each axis. Firstly forming low-density sheet-like elements, further collapse pushes the matter in long string-like filamentary patterns, after which the matter collapses in each direction and forms high-density clusters. Regional grids of filaments also compile sheets in $2-$dimensions.

Data that emerged from deep redshift surveys, such as the Center for Astrophysics (CfA) redshift survey \citep{1983sgrd.book.....H} and The Queen Mary and Westfield College, Durham, Oxford, and Toronto (QDOT) all-sky IRAS galaxy redshift survey \citep{1999MNRAS.308..897L} showed that the Universe can be described as having a sponge-like structure. \citet{1986ApJ...302L...1D} studied the data of $1099$ galaxies from the CfA survey. The analysis revealed almost empty bubbles (diameters between $25 - 50 h^{-1} \mathrm{Mpc}$) and elongated regions of galaxies bordering them. The results challenged the then-accepted models for structure formation in the Universe.
The analysis of \citet{1986ApJ...306..341G} showed that low-density regions are connected by tunnels and high-density regions are connected by bridges. The creation of this sponge-like structure of the Universe can be explained by the Gaussian random-phase initial conditions. \citet{1992MNRAS.256..477M} compared the QDOT survey topology with an artificial QDOT survey (a corresponding N-body simulation using the standard cold dark matter model) to analyse whether the observed distribution of galaxies is the result of the gravitational growth of the initial Gaussian field of density fluctuations. On length scales of $10-50 h^{-1} \mathrm{Mpc,}$ the data topology corresponded well with the distribution that evolved from the Gaussian field of density fluctuations. Both maps showed that high-density regions are joined by bridges and low-density regions have a sponge-like interconnection. \citet{1994ApJ...420..525V} analysed the CfA survey and clearly showed the Universe's sponge-like topological nature of high-density regions and empty areas. 
Galaxy redhshift surveys that were orders of magnitudes larger were made possible in this century. Some of the best-known surveys are the 2dF Galaxy Redshift Survey (2dFGRS) \citep{2001MNRAS.328.1039C} and the Sloan Digital Sky Survey (SDSS) \citep{2000SDSS}. \citet{2009ApJ...695L..45G} again revealed a sponge-like topology of the luminous red galaxies in the Sloan Digital Sky Survey $3-$dimensional map. This topology is expected from the Gaussian random-phase initial conditions.

As the evolution of the large-scale structure is driven by density fluctuations, it is itself controlled by gravity~\citep{peebles1980large}. Gravitational instability is responsible for the growing contrast in the matter distribution in the Universe~\citep{noam2017tempel}. The complex structure and evolution of the web plays an important role in understanding the elements residing in the structures \citep{2011A&A...529A..53T, 2017MNRAS.464..666B, 2018MNRAS.481..414G}.

To study the galaxy distribution morphology in the large-scale structure of the Universe, galaxies are often viewed as points in 3-dimensional Euclidean space~\citep{martinez2001statistics}. The Cartesian coordinates of the galaxies are the point locations in space, and their other properties are viewed as marks. Corollary for the observed data of galaxies in a survey region can be viewed as a realisation of a marked point process in a compact set. This opens a pathway to use spatial statistics to investigate the distribution of these galaxies. 

The mathematical framework of the bisous model~\citep{TempEtAl14}, developed to estimate filamentary patterns from observed spectroscopic galaxy data, reveals an intrinsic network. 
In this study we analyse whether these detected galaxy filaments can also be seen by the SDSS photometric redshift galaxies dataset compiled in~\citet{beck2016photometric}. Because the photometric redshift galaxy datasets will be more accurate and extend to higher redshifts in the coming galaxy redshift surveys such as the Javalambre Physics of the Accelerating Universe Astrophysical Survey (J-PAS) ~\citep{benitez2014j}, the use of these galaxies in cosmic web structure modelling becomes more prominent.

In this study we aim to show whether these additional galaxies carry information about the filamentary network and either prove or refute the estimated parameter for filament thickness reported in~\citet{TempEtAl14}. 
By applying summary statistics to analyse the possible correlation between the investigated sets of objects, it is possible to determine whether the photometric redshift galaxies are positively associated with the filamentary network of the cosmic web.
Another aspect of interest is whether these galaxies show a sign of being inside these filamentary structures. By calculating physical distances from the photometric galaxies to the filamentary network axes, we estimate the filament thickness, where these galaxies reside, and we can compare it with the radius obtained in~\citet{TempEtAl14} by the bisous model from the spectroscopic galaxies.

We wish to broaden the view for the algorithms that detect the filamentary spine and show that photometric redshift galaxies together with spectroscopic galaxies can be used to detect the filamentary pattern in the cosmic web of galaxies. The photometric redshift galaxy catalogues describe objects that are more distant and/or fainter than the spectroscopic galaxies datasets, thus broadening the space to detect the cosmic web pattern. Photometric galaxies would increase the number density of galaxies per volume of space and thus benefit the filamentary pattern detection process.
If these galaxies are located inside or close to the filamentary spines, this would mean that their contribution to filamentary pattern modelling could be significant. The bisous model extracts the filamentary pattern from observational galaxy data. With the growth of the galaxy data, the filamentary pattern recognition becomes more precise, and a more complete structure will be obtained. 

The structure of the paper is as follows. In Sect.~\ref{Data} we characterise the datasets and their subsets. In Sect.~\ref{mathematicalframework} we describe the mathematical frameworks we used to answer the questions at hand. In Sect.~\ref{results} we present the results on the sphere and in $3-$dimensional Euclidean space. In Sect.~\ref{conclusion} the obtained results are discussed and conclusions are drawn. 

\section{Data}\label{Data}
The datasets we used in this work are the following: the Sloan Digital Sky Survey~\citep{2000SDSS} Data Release $12$ of spectroscopic galaxies used to derive the filamentary web, the filamentary pattern catalogue detected from the spatial distribution of the spectroscopic galaxies with the method described in~\citet{TempEtAl14}, and a catalogue of photometric redshift galaxies~\citep{beck2016photometric}. This last catalogue holds information about galaxies that were not used to detect the filamentary web. The interest is to see whether these photometric galaxies possess information about the filamentary network that is detected from the spatial distribution of spectroscopic galaxies. Catalogues describing galaxies and defined filamentary structures of the cosmic web are well defined either in the sphere or in the $3-$dimensional Euclidean space. The analysis is therefore done both in a region of a sphere $\mathbb{S}^{2}$ and in the $3-$dimensional Euclidean space $\mathbb{R}^{3}$.

\begin{figure}
   \centering
   \includegraphics{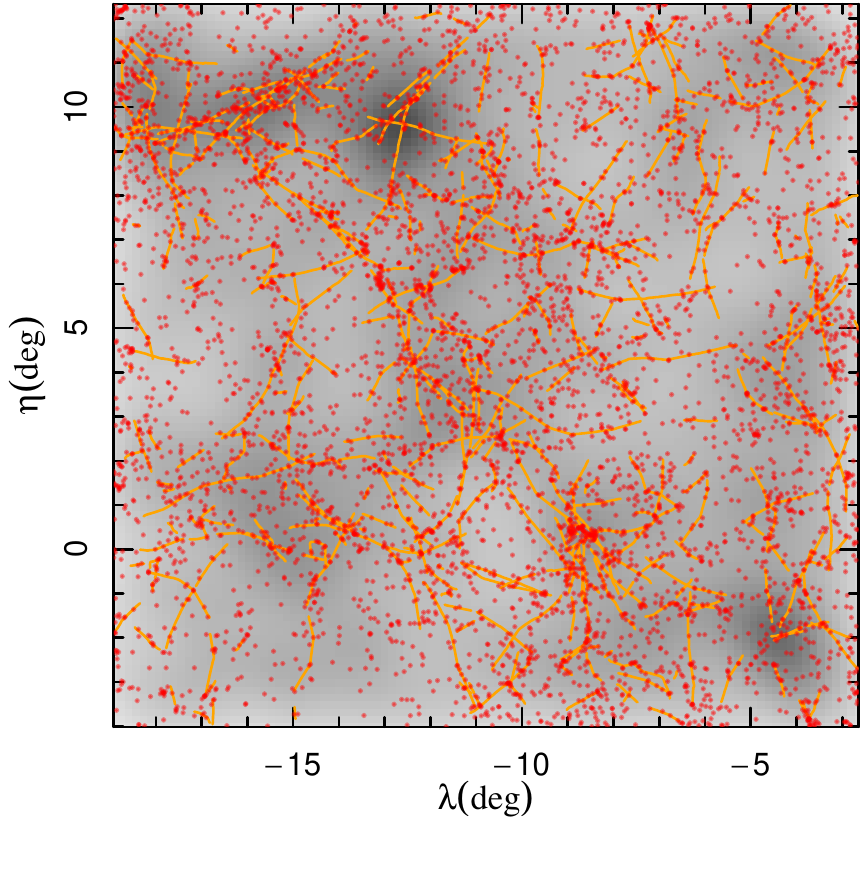} 
     \caption{Visualisation of the datasets in spherical sky-coordinates: kernel-smoothed 2-dimensional density from the distribution of photometric galaxies on the viewed region of the sphere (grey background density), filamentary spines (orange spines), and spectroscopic galaxies (red dots). Darker grey denotes a higher density of photometric galaxies. All the drawn objects are located in the distance range of $400-600$~Mpc.} 
         \label{Photoz200800_Others200800}
   \end{figure}

Figure~\ref{Photoz200800_Others200800} describes the datasets of filamentary spines, spectroscopic galaxies, and the photometric galaxies on the sphere. The objects are drawn using their  longitude $\lambda$ (deg) and latitude $\eta$ (deg) to produce this plot. The grey background $2-$dimensional density was created by smoothing the photometric galaxies distribution with a kernel. The darker grey color denotes the higher density of photometric galaxies in that region. Figure \ref{Photoz200800_Others200800} illustrates that the higher density regions of photometric galaxies coincide quite well with the higher number of detected filamentary spines and spectroscopic galaxies. As expected, the filamentary spines (orange lines) tend to follow the distribution of the spectroscopic galaxies (red dots). 

 This section continues to describe the analysed datasets, mappings, and catalogue subsamples. The first subsection gives an overview of the dataset of spectroscopic galaxies. The second subsection describes the filamentary structure data. The third subsection introduces the photometric redshift galaxy data. Lastly, we characterise the subsamples taken from all the datasets of interest. 

\subsection{Spectroscopic galaxy data}\label{SpectroscopGal} 
 
The spectroscopic galaxy dataset used in this paper was obtained from \textit{CosmoDB}\footnote{Catalogue collection compiled and published by researchers at the Tartu University Tartu Observatory; www.cosmodb.to.ee}, which is a cosmology-related catalogue collection.
The galaxies are taken from the main contiguous area of the SDSS ($7966$ sq deg) with the redshift limit of $z=0.2$ and $r$-magnitude limit of $17.77$. In this catalogue, the Finger-of-God effect of galaxy groups or clusters is suppressed with the friends-of-friends (FoF) algorithm; see~\citet{2017A&A...602A.100T} for a detailed description. A brief description is given in the following.

Large fractions of galaxies tend to reside in galaxy groups. The peculiar velocities of galaxies residing in these groups or clusters elongates the compact structures along the line of sight in redshift space, which seemingly distorts the galaxy distribution.  
This effect was first described by~\citet{1972MNRAS.156P...1J}
and was later introduced by~\citet{FingerOfGodTully} as the Finger-of-God effect. The detection of groups from the large-scale data and the suppression of the distorted structures is of great importance because it influences the overall distribution of galaxies and their allocation to different types of environments. The Finger-of-God effect is eliminated for the galaxies forming groups in the SDSS spectroscopic galaxies dataset by the FoF algorithm, which was first described for a cosmological context by~\citet{TurnerFOF}. A catalogue of groups detected from the SDSS spectroscopic galaxy data by the FoF algorithm is described in~\citet{TempelGroups2012}. These galaxy groups are detected from the distribution of galaxies that are bound into groups using the algorithm with a certain linking length. \citet{2014A&A...566A...1T} described the linking length thoroughly. In brief, it can be described as the maximum distance from an observed galaxy to determine its neighbours in the galaxy group-finding algorithm. 

The dataset describes $584~449$ spectroscopic redshift galaxies. These galaxies describe a $3-$dimensional spatial point pattern from which the filamentary structure was detected by the method described in~\citet{TempEtAl14}, see Sect.~\ref{Filamentarystructuredata}. 
The spectroscopic redshift galaxies are further viewed as a set of points $X_{\mathrm{spec}}$, which are described in the $3-$dimensional Euclidean space by their Cartesian coordinates $(x,y,z)$. Each galaxy is also attributed spherical coordinates of latitude $\eta$ and longitude $\lambda$ from the catalogue.
An estimate for the spectroscopic galaxy distance from the observer is obtained from the catalogue. Sampling of the spectroscopic galaxies according to their distance from the observer was done and is further discussed in Sect.~\ref{tomography}. 

\subsection{Filamentary structure data}\label{Filamentarystructuredata}

The catalogue of the filamentary network was obtained from the spatial distribution of spectroscopic redshift galaxies (Sect.~\ref{SpectroscopGal}). The filaments were detected by applying a marked point process to the galaxy field~\citep{TempEtAl14}. The description of this process, the bisous model, and more general references related to the theory of marked point processes, can be found in~\citet{Stoica2010} and~\citet{TempEtAl14,2016A&C....16...17T}.

The bisous model estimates the filaments of the cosmic web by a random configuration of connected and aligned cylinders. This configuration is obtained through maximising a probability density that governs the position and interaction of these cylinders. The maximisation is done through the simulated annealing algorithm. The stochastic nature of the optimisation procedure enables this method with several outputs:
\begin{itemize}
\item[•] configuration of aligned and connected cylinders;
\item[•] visit map field: assigns to each point of the observed domain the probability to be "touched" by the filamentary network;
\item[•] spines: ridge lines of the highest values within the visit map field.
\end{itemize}

We used a set of $46\,403$ spines. The set of spines forms a random set (see Sect.~\ref{mathematicalframework}) of objects $Y_{\mathrm{fil}}$ in the $3-$dimensional Euclidean space. A point in $\mathbb{R}^{3}$ was mapped to the spherical coordinates latitude $\eta$ and longitude $\lambda$ using
\begin{eqnarray}
\eta & = & 2 \arctan{\left(\frac{y}{\sqrt{x^{2}+y^{2}}}-x\right)} ,  \nonumber \\ 
\lambda & = & \arcsin{\left(\frac{z}{\sqrt{(x^{2}+y^{2}+z^{2})}}\right)}. \label{tospheric}
\end{eqnarray}
The mean orientation towards the line of sight of a spine $Y$ is defined as
\begin{equation}
\cos(\bar{\alpha}(Y)) = \frac{1}{l(Y)}\int\limits_{Y} \boldsymbol{\omega}_{y} \cdot \mathbf{\bar{O}}y \mathrm{d}y\label{alpha}
,\end{equation}
with $\boldsymbol{\omega}_y$ the orientation vector of the filamentary network in point $y$ of the spine $Y$ and $\bar{\mathbf{O}}y$ the line-of-sight vector at point $y$. Value $\bar{\alpha}(Y)$ represents the average angle that the spine $Y$ has corresponding to the line of sight. 
The mean distance from the spine $Y$ to the observer $O$ is defined as
\begin{equation}
\bar{d}(O,Y) = \frac{1}{l(Y)}\int\limits_{Y}d(O,y)\mathrm{d}y
\label{spinedistance}
,\end{equation} 
while the integral is computed along the considered spine, $l(Y)$ is the associated spine length, and $d(O,y)$ is the distance from the observer $O$ to point $y$ of the spine $Y$.

\subsection{Photometric galaxy data} \label{PhotometricData}
The SDSS Data Release $12$ offers catalogues of photometric redshift galaxies. The dataset used in this analysis and the method used to estimate the photometric redshifts of the galaxies are presented in \citet{beck2016photometric}. 
Briefly, the photometric redshift and its error were estimated with an empirical method called local linear regression.
A detailed description of the properties and a description of the photometric redshift derivation is given in \citet{beck2016photometric}.

For the current analysis, we chose photometric redshift galaxies in the same sky region as the filamentary structure at the maximum redshift $z_{\mathrm{photo}}=0.25$ , with the photometric redshift error estimate $\delta z_{\mathrm{photo}}\leq 0.05$. 
The dataset of photometric galaxies was limited in this manner to obtain the set of galaxies with the best redshift estimates, to have a reasonable number of galaxies in the analysed set, and so that the redshift limit coincided with the limit of the other dataset redshifts. The final dataset contained $2~893~018$ galaxies. Their locations are defined on the sphere by their radian latitude $\eta$ and longitude $\lambda$. These photometric galaxies are viewed as a configuration of points $X_{\mathrm{photo}}$ in a sphere $\mathbb{S}^{2}$. 

The error estimates for the photometric redshifts of the galaxies $\delta z_{\mathrm{photo}}$ reach up to $0.05$. This makes the use of the redshifts in describing the galaxy location rather uncertain. 
This results in the absence of an accurate $3-$dimensional coordinate system for the galaxy positions. 
These  quite uncertain photometric redshift distance estimates were not used to calculate the distances between the observed sets. The photometric galaxy redshift estimates were used to sample the data into subsequent distance ranges, however, which we discuss in Sect.~\ref{tomography}. 

\subsection{Subsamples}\label{tomography}

To test possible systematic effects of differently oriented filamentary spines, the spines were sampled dependent on their angle $\bar{\alpha}$ corresponding to the line of sight. The tested systematics are described below. 
To determine whether the photometric redshifts are of use and might carry information about the filamentary spine locations, the photometric galaxies were distributed into distance slices dependent on their redshift. To sample the spectroscopic galaxies, photometric galaxies, and filamentary data dependent on their distance, we used the estimated distances to the galaxies from the observer. 
  
\begin{figure}
   \centering
   \includegraphics{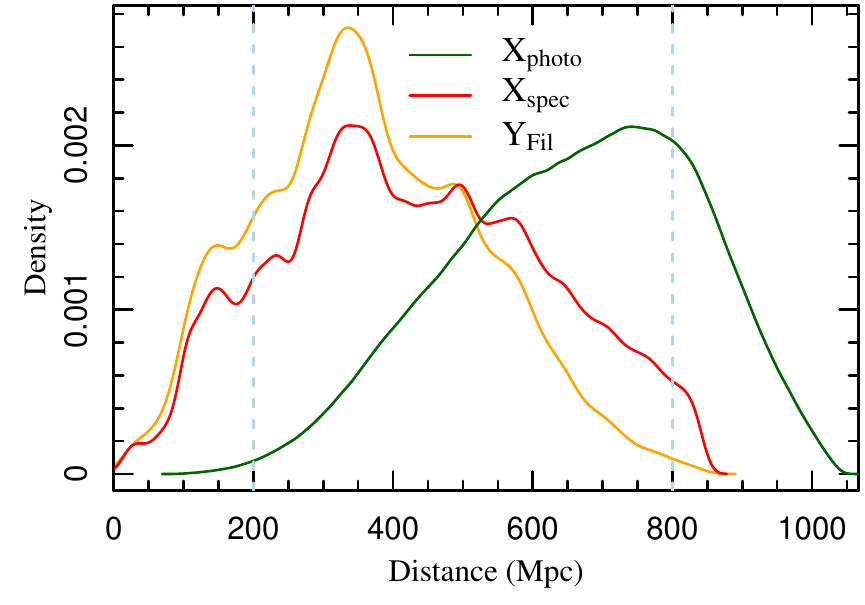}
     \caption{Density distribution of the spectroscopic redshift galaxies $X_{\mathrm{spec}}$ (red line), all filamentary spines $Y_{\mathrm{fil_{all}}}$ (orange line), and photometric redshift galaxies $X_{\mathrm{photo}}$ (green line) distances from the observer of all corresponding objects in the described datasets. The distance distribution is normalised individually for each dataset. See Table~\ref{AlphaSets} for the actual number of objects. The dashed light blue lines denote the $200$ to $800$~Mpc distance range used here.}
         \label{AllFilSpecPhotoDistDB}
   \end{figure}

Figure~\ref{AllFilSpecPhotoDistDB} presents the density distribution of the distances from the observer of photometric redshift galaxies compared to the density distribution of spectroscopic redshift galaxies and filamentary spine distances from the observer. It shows the difference in the number of objects in the observed sets on the distance scale. There are about 4.5 times more observed galaxies in the photometric galaxy dataset than in the spectroscopic galaxy sample. Photometric galaxies complement the sample mostly farther away, where the spectroscopic sample is more incomplete. They might add much information to the galaxy point data by filling in the possible gaps in the cosmic web network and creating a more precise vision of the large-scale structure.

Figure \ref{AllFilSpecPhotoDistDB} shows that spectroscopic galaxies and filamentary spines cover the same area on the distance scales roughly equally. This is highly expected because the filamentary spines are detected from the spatial distribution of these spectroscopic galaxies. An overwhelming abundance of photometric galaxies at greater distances from the observer is contrary to the case of spectroscopic galaxies. Because only a few spectroscopic galaxies are observed at greater distances, the filamentary spine finder starts to fail beginning at a distance of $600$~Mpc. The abundance of photometric galaxies greater distances might remedy this effect, 
although in the currently available catalogues of photometric galaxies, the distances are only roughly known, which makes the use of them in galaxy pattern detection from observational data more complex. This effect will be very weak in the upcoming photometric  redshift galaxy surveys such as the J-PAS~\citep{benitez2014j}, where photometric redshifts of galaxies up to redshift $1$ will be known with unprecedented accuracy. This provides even more motivation to use galaxy photometric redshift data in future cosmic web modelling. 

Figure~\ref{AllFilSpecPhotoDistDB} also exhibits the low density of photometric redshift galaxies below $200$~Mpc and the low density of filamentary spines above the $800$~Mpc threshold. We therefore limit for the entire analysis all of the datasets to between $200$~Mpc and $800$~Mpc. This distance range is highlighted in the figure with light blue dashed lines.  
The spines were filtered using their average distance of the filamentary spine axis (Eq.~\ref{spinedistance}), and the spectroscopic galaxies were filtered by their distance estimates in the catalogue. 
To filter the photometric galaxies into observed samples, their photometric redshift distance estimates from the catalogue were used. The number of objects in each of the datasets fulfilling these conditions is given in Table~\ref{AlphaSets}.
 
Fig.~\ref{AllFilSpecPhotoDistDB} also shows the distribution of average distances of the filamentary spine axes in $200$~Mpc wide distance slices in the range of $200$ to $800$~Mpc. The innermost range of $400-600$~Mpc is most uniformly distributed for all subsets. Different distance ranges are not equally well described by filamentary data or the galaxy datasets.
Figure \ref{AllFilSpecPhotoDistDB} clearly shows that when the number of spectroscopic galaxies decreases, the number of photometric galaxies increases. The catalogue of filamentary spines and the dataset of photometric galaxies was divided into subsequent distance ranges from the observer. This was done to analyse the following three aspects: Are the results affected by the evolution of the number of galaxies in the sets over the distance scale? How much extra information do the photometric redshifts hold about the locations of the filamentary spines? What is the origin of the possible signal? In this way, we analysed the filamentary spine catalogue and the photometric redshift galaxy catalogue in $200$~Mpc wide distance ranges from $200$~Mpc until $800$~Mpc. The width of $200$~Mpc was chosen because the number of filamentary spines, spectroscopic galaxies, and photometric galaxies is sufficiently high in these ranges. The number of objects and notation of the sets in the subsequent $200$~Mpc wide distance ranges is given in Table \ref{AlphaSets}. 

The possible correlation was calculated between photometric galaxies and filamentary spines in the sphere $\mathbb{S}^{2}$. For this, the set of points defining the filamentary spine axis was mapped to the sphere. This distorts the filamentary spine objects, which are not perpendicular to the line of sight. The distortion is partly eliminated when we sample the filaments to be fairly perpendicular to the line of sight. In addition, this created an opportunity to determine whether the orientation of the axis affects the clustering or alters the physical distances between the observed sets. This effect might occur because the compression of the Finger-of-God effect with the FoF algorithm might compress filamentary spines along the line of sight.

\begin{figure}
   \centering
   \includegraphics{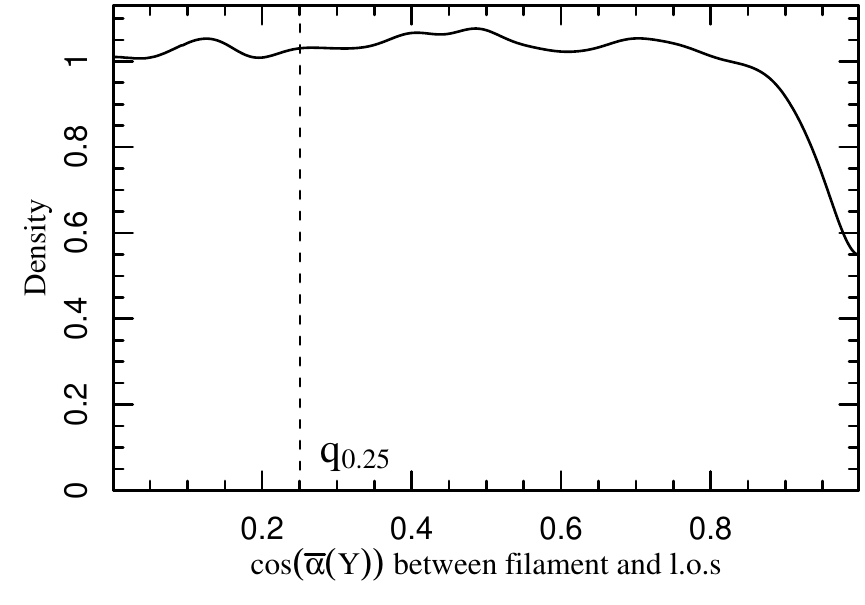}
     \caption{Smoothed density distribution of the cosine of the mean orientation of spines $\cos(\bar{\alpha}(Y))$ of the filamentary spines whose average distance is in the range from $200$~Mpc to $800$~Mpc. The dashed black line denotes the $\mathrm{q}_{0.25}$ quantile of the density distribution, which is used to select subsamples of filaments that are perpendicular to the line of sight.}
              \label{AverageCosine_PerFil_Degrees}
   \end{figure}

To investigate whether any such systematic effects might arise from the different orientations of the filamentary spines, the filamentary spine data were sampled according to the angle $\bar{\alpha}$ (Eq. \ref{alpha}).
Figure~\ref{AverageCosine_PerFil_Degrees} represents the smoothed density distribution of the cosine of the average angle $\cos(\bar{\alpha})$ of the studied filamentary spines $Y_{\mathrm{fil}}$.
The value $\cos(\bar{\alpha}) = 0 $ represents the instance when the filamentary spine is at the angle of $90 ^\circ$ with the line of sight, which means that they are perpendicular. To subset all of the filamentary spines into those perpendicular and not perpendicular to the line of sight, we used the $\alpha-$quantiles. The $0.25-$quantile was chosen for this, and with respect to the dataset we used, it roughly results in the fact that all filaments whose average angle $\bar{\alpha}$ is in the angle range of $78.22^\circ$ to $90^\circ$ were viewed as perpendicular. The curve in Fig.~\ref{AverageCosine_PerFil_Degrees} shows fewer filamentary spines parallel ($\bar{\alpha}=0 ^\circ$) to the line of sight than are completely perpendicular ($\bar{\alpha}=90 ^\circ$). 

Table \ref{AlphaSets} contains all the subsamples taken from the datasets in the previously described fashion. It consists of the subsample name, notation, numbers, and the set conditions for sampling. 

\begin{table*}
\centering
\caption{Sampling the catalogue of filamentary spines and the two galaxy datasets. Sampling is conditioned on datasets in the distance range of $200-800$~Mpc. The notation $N$ represents the counting measure, the distance is the object distance from the observer, and $\bar{\alpha}$ is the average angle of the filamentary spine corresponding to the lines of sight (Eq. (\ref{alpha})). Perpendicular filaments and non-perpendicular filaments are the spines that are or are not perpendicular towards the line of sight, respectively.}\label{AlphaSets}
\begin{tabular}{ p{8.0cm} p{2cm} p{2.8cm}p{1.4cm} p{1.6cm}  }
 \hline
  \hline
Subsample & Distance (Mpc) & $\bar{\alpha}$ (deg) & $N$ & Sample notation  \\[1pt]
\hline
non-perpendicular filaments & $200-800$ & $0^\circ \leq \overline{\alpha} < 78.22^\circ $
& $30~639$ &$\mathrm{fil_{notperp}}$   \\ 
perpendicular filaments & $200-800$ & $78.22^\circ \leq \overline{\alpha} \leq 90 ^\circ$ 
& $ 8~063$ & $\mathrm{fil_{perp}}$ \\ 
filaments & $200-800$ & all $\overline{\alpha}$ & $38~702$ & $\mathrm{fil_{all}}$ \\
filaments & $200-400$  & all $\overline{\alpha}$ & $20~367$ & $\mathrm{fil}_{200-400}$ \\
filaments & $400-600$  & all $\overline{\alpha}$ & $14~598$ & $\mathrm{fil}_{400-600}$ \\
filaments & $600-800$  & all $\overline{\alpha}$ & $3~737$ & $\mathrm{fil}_{600-800}$ \\
spectroscopic galaxies & $200-800$ & - & $497~611$ & $\mathrm{spec_{all}}$ \\
spectroscopic galaxies inside filaments & $200-800$ & all $\overline{\alpha}$ & $139~051$ & $\mathrm{specfil_{all}}$ \\
spectroscopic galaxies inside perpendicular filaments & $200-800$ &  $78.22^\circ \leq \overline{\alpha} \leq 90 ^\circ$ & $28~486$ & $\mathrm{specfil_{perp}}$ \\
spectroscopic galaxies inside non-perpendicular filaments & $200-800$ &  $0^\circ \leq \overline{\alpha} < 78.22^\circ$ & $110~565$ & $\mathrm{specfil_{notperp}}$ \\
photometric galaxies & $200-800$  & - & $2~198~702$ & $\mathrm{photo_{all}}$ \\ 
masked photometric galaxies\tablefootmark{a} & $200-800$  & - & $570~083$ & $\mathrm{photo_{mask}}$\\
photometric galaxies & $200-400$  & - & $236~850$ & $\mathrm{photo}_{200-400}$\\        
photometric galaxies & $400-600$  & - & $802~970$ & $\mathrm{photo}_{400-600}$\\                                
photometric galaxies & $600-800$  & - & $1~158~882$ & $\mathrm{photo}_{600-800}$ \\                      
 \hline
 \end{tabular}
 \tablefoot{
\tablefoottext{a}{Masked-out photometric galaxies in the observed region of the sphere dependent on the locations of spectroscopic galaxies. A detailed description is given in Sect.~\ref{Bivariateresults}.}
}
\end{table*}
 
\section{Mathematical framework}\label{mathematicalframework}

The data used in this paper are particularly adequate for being analysed with spatial statistics tools~\citep{martinez2001statistics,baddeley2015spatial}. Three datasets were investigated: spectroscopic galaxies, photometric galaxies, and filamentary spines. It is the classical assumption to consider the first two sets as point-processes realisations. Because of its complexity, the third set is assumed to be the realisation of a random set. This is a more general stochastic process than point processes. From a more intuitive perspective, point processes consider random configurations of points, while random sets are used for random configurations of geometrical objects. Any point process is a random set, while the reverse does not always hold. For a thorough presentation of both these mathematical objects, we refer to \citet{ChiuEtAl13} and the references therein.

This section starts with a brief presentation of point processes, and it recalls the definition of the Poisson point process, which is maybe the most important point process because of its remarkable properties. Next, the section presents summary statistics that allow characterising an observed random set with respect to the related Poisson process~\citep{baddeley2015spatial}. Furthermore, the summary statistics may be used to asses a type of association when it is applied to a pattern of different types~\citep{foxall2002nonparametric,kleinschroth2016effects}. Finally, the section ends by presenting some estimation techniques and computational aspects.

\subsection{Point processes} 
Point processes are probabilistic models allowing the study of a random configuration of points in an observation window. These tools have been successfully used for the statistical analysis of astronomical datasets for more than half a century~\citep{martinez2001statistics}. Rigourous and detailed mathematical presentations may be found in the monographs of~\cite{Lies00},~\citet{MollWaag04},~\citet{illian2008statistical},~\cite{ChiuEtAl13}, and~\citet{baddeley2015spatial}. The current presentation follows~\citet{StoicaMarkedPoints} and~\citet{baddeley2007spatial}.

Let $W \subset \mathbb{R}^{3}$ be a compact set, the observed finite region of the Universe.
Let $\nu$ be the Lebesgue measure in $\mathbb{R}^{3}$ and $(W, \mathcal{B}_{W}, \nu)$ the natural restriction to $W$ of $(\mathbb{R}^{3}, \mathcal{B}, \nu)$,
with $\mathcal{B}$ being the associate Borel $\sigma-$algebra. 
$W_{n}$ is the set of all unordered configurations $\mathbf{w}=\{w_{1},w_{2},...,w_{n}\}$, which consist of $n$ not necessarily distinct points $w_{i} \in W$, for $n \in \mathbb{N}$. We consider the configuration space given by $\Omega_{W}=\bigcup_{0}^{\infty} W_{n}$ equipped with the $\sigma -$ algebra $\mathcal{F}_{W}$ generated by the mappings $\{w_{1},...,w_{n}\} \rightarrow \sum_{i=1}^{n}\mathbbm{1} \{w_{i} \in B\}$ that count the number of Borel sets $B \in \mathcal{B}_{W}$. 
A point process on $W$ is a measurable map from a probability space $(\Omega_{W}, \mathcal{F}_{W})$. 
In other words, a point process is a random configuration of the points lying in $W$.

The Poisson point process in $W$ is built as follows. First, let $\beta : W \rightarrow ]0, + \infty[$ be the intensity function. Then, a Poisson process is defined by the following two properties:
\begin{enumerate}[i)]
\item for any bounded set $B$, $N(B)$, the number of points in $B$ is a Poisson random variable with mean $\Lambda(B)= \int_{B} \beta(x)\mathrm{d}\nu(x)$,
\item for any $B_{1},...,B_{m}$ disjoint bounded sets, then point counts $N(B_{1}), ..., N(B_{m})$ are independent random variables.
\end{enumerate}
If the Poisson point process has a constant intensity $\beta=const$, then the process is said to be stationary or homogeneous.

\subsection{Summary statistics}
\label{J-function}
An exploratory analysis of observed point patterns is typically performed using summary statistics~\citep{MollWaag04,Lies00}. 
For a stationary point process $X$ of intensity $\beta >0$, some of the most frequently used summary statistics are
\begin{itemize}
\item the nearest-neighbour distance distribution function
\begin{equation}
G(r) = \PP_{o}^{!}(N(b(o,r) > 0))
\label{eqNearestNeighbour}
,\end{equation}
with $\PP_{o}^{!}$ the reduced Palm distribution and $N(b(o,r))$ the number of points in a ball of radius $r$ centred at the origin $o$. The origin $o$ is a point of the considered stationary process that is not{\it } counted, 
\item the empty space function
\begin{equation}
F(r) = \PP(X \cap b(o,r) \neq \emptyset)
\label{eqEmptySpace}
,\end{equation}
with $\PP$ the distribution of $X$,
\item the $J$ function that compares nearest-neighbour distances to empty distances,
\begin{equation}
J(r) = \frac{1-G(r)}{1-F(r)}
\label{eqJFunction}
,\end{equation}
for all $r > 0$ such that $F(r) < 1$.
\end{itemize}
These summary statistics have exact formulas for the stationary Poisson point process with intensity $\beta$~:
\begin{eqnarray}
F(r) & = &  1 - \exp[-\beta \pi r^2], \nonumber\\
G(r) & = & F(r), \nonumber \\
J(r) & = & 1.
\end{eqnarray}

For a given point pattern, the estimators of these statistics give indications about how far from the Poisson point process the considered pattern is. For the $G$ statistics, a lower value of the observed summary than the theoretical Poisson statistics suggest regularity or repulsion for the pattern, while the higher value suggests clustering. For the $F$ and $J$ statistics, it is the other way around: higher observed values than the theoretical Poisson indicate repulsion, while the lower values recommend clustering.

To give more reliable answers to this type of questions, an envelope test can be made. This test simulates a Poisson point process with intensity estimated from the observed pattern, and for each simulated pattern, the summary statistics are computed. At the end of the simulation, Monte Carlo confidence envelopes are created. This means that if the observed summary statistics belongs to the envelope, the null hypothesis that the observed point pattern is the realisation of a Poisson process is not rejected, hence the point pattern exhibits a completely random structure. For a thorough presentation of the point-process summary statistics and of the envelope tests based on them, we 
recommend~\citet{Lies00}, \citet{MollWaag04}, \citet{illian2008statistical} and \citet{MyllEtAl17} and the references therein.

Mathematical developments allow extending these functions to more general patterns~
\citep{lieshout1996nonparametric,
LiesBadd99,foxall2002nonparametric,ChiuEtAl13,2016arXiv160402302V}. In the following, we introduce the bivariate $J$ function.

We consider $X$ a point process and $Y$ a random set, and we assume that $(X,Y)$ is jointly stationary~\citep{foxall2002nonparametric}. This means that the distributions of the bivariate process $(X,Y)$ and of the bivariate process $(X+e, Y+e)$ are identical for any transition vector $e \in \mathbb{R}^{3}$.

The definition of the empty space function in Eq.~\eqref{eqEmptySpace} can be extended to random sets $Y$ in a straightforward way by replacing $X$ with $Y$ in the formula.
The nearest-neighbour function~\eqref{eqNearestNeighbour} becomes $G_{X,Y}$ , the distribution function of the distance from a typical point of the point process $X$ to the nearest objects of the random set $Y$:
\begin{equation}
G_{X,Y}(r) = \PP^{!}_{o}(b(o,r) \cap Y \neq \emptyset),
\end{equation}
where $\PP^{!}_{o}$ represents the reduced Palm distribution at an arbitrary point $o$ of the point process $X$.

Similarly to Eq.~\eqref{eqJFunction}, the bivariate $J-$function is given by
\begin{equation}
J_{X,Y}(r)=\frac{1-G_{X,Y}(r)}{1-F_{Y}(r)}
\label{eqBiVariateJ}
\end{equation}
for all $r \geq 0$ such that $F_{Y}(r) < 1$.

The $J_{X,Y}-$function can be interpreted as a measure of association between $X$ and $Y$~\citep{foxall2002nonparametric}. If $X$ and $Y$ are independent, then $G_{X,Y} = F_{Y}$ and $J_{X,Y} = 1$. Values of $J_{X,Y}$ close to $1$ therefore suggest independence, values higher than $1$ suggest negative association or ``repulsion'', and values lower than $1$ suggest positive association or ``clustering''.

\subsection{Estimation}
\label{Estimation}
The previous summary statistics should be estimated from a pattern that is observed in a bounded window only. This introduces edge effects. Several strategies are available to overcome this problem~\citep{baddeley2015spatial}. Here, we adopted the border-corrected estimation as in~\citet{lieshout1996nonparametric} and~\citet{foxall2002nonparametric}.

We consider the shortest distances from a point $w \in W$ to a subset $A \subset W,$ 
\begin{equation*}
d(w, A)=\mathrm{inf}_{a \in A} \|  w-a \| . 
\label{shortestdistance}
\end{equation*}
The border-corrected estimator for the empty space function is
\begin{equation}
\widehat{F}_{Y}(r)= \frac{\sum_{i} \mathbbm{1}\{d(w_{i}, W^{c}) \geq r\} \mathbbm{1} \{d(w_{i}, Y) \leq r\}}{\sum_{i} \mathbbm{1}\{d(w_{i}, W^{c}) \geq r\} }
\label{F}
,\end{equation}
with $W^{c}$ the border of $W$ and $\{w_i,i=1,2,\ldots\}$ a finite family of arbitrary points in $W$. The most common choice for the set of arbitrary points is a regular grid.

The estimator of the nearest-neighbour distribution from any point in $X$ to the random set $Y$ is
\begin{equation}
\widehat{G}_{X,Y}(r)= \frac{\sum_{i} \mathbbm{1}\{d(x_{i}, W^{c}) \geq r\} \mathbbm{1} \{d(x_{i}, Y) \leq r\}}{\sum_{i} \mathbbm{1}\{d(x_{i}, W^{c}) \geq r\} },
\label{G}
\end{equation}
where $\{x_i,i=1,\ldots\}$ is the observed finite-point configuration of $X$.

The bivariate $J$ function is obtained by plugging in~Eqs. \eqref{F} and~\eqref{G} in Eq.~\eqref{eqBiVariateJ}:
\begin{equation}
\widehat{J}_{X,Y}(r) = \frac{1- \widehat{G}_{X,Y}(r)}{1-\widehat{F}_{Y}(r)} .
\end{equation}
This estimator is consistent and asymptotically normal, while $W$ increases and $r$ is fixed. An approximation of its variance is given in~\cite{foxall2002nonparametric}.

   \begin{figure*}
   \centering
   \includegraphics[width=\textwidth]{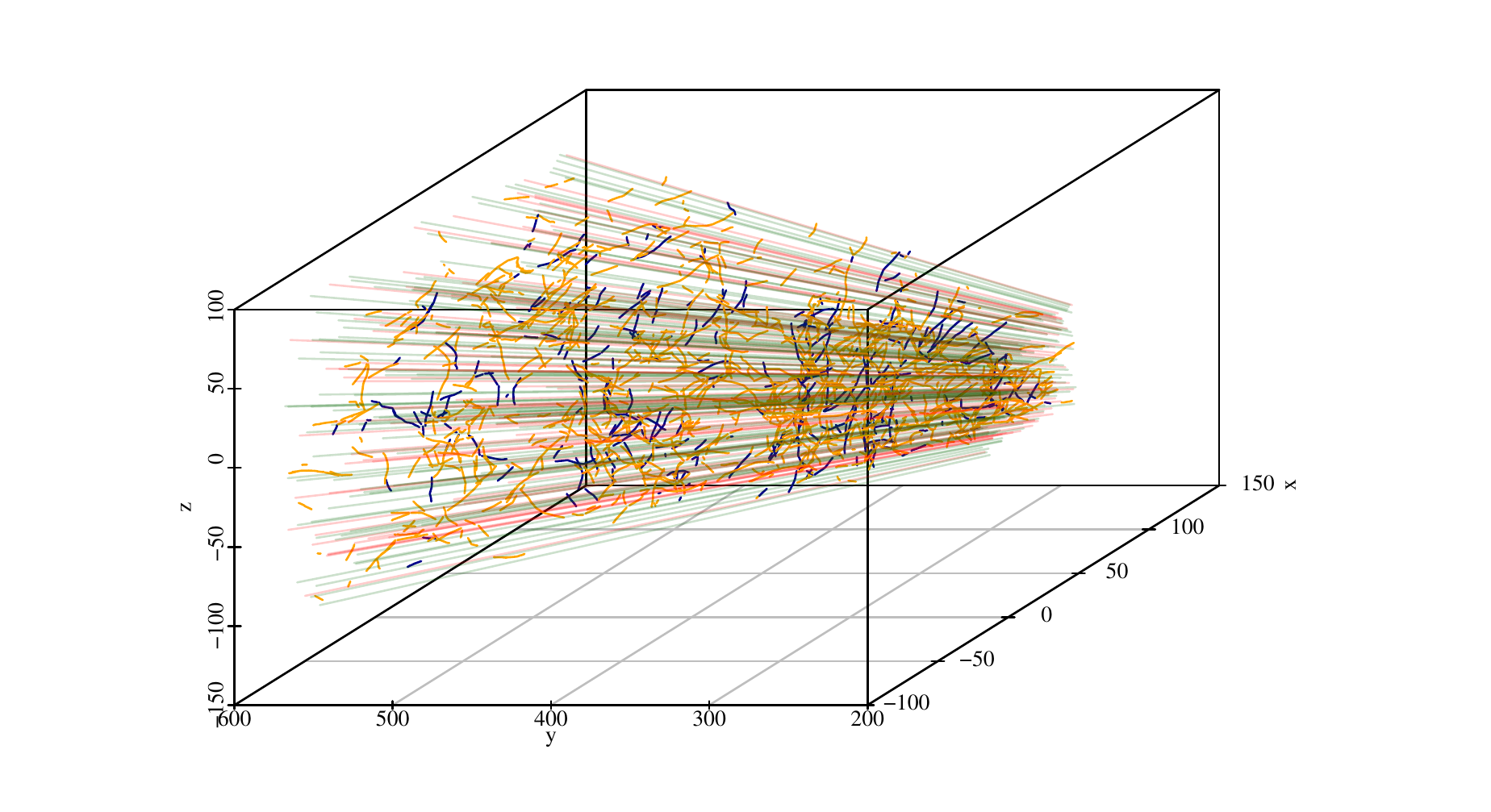}
     \caption{Visualisation of the datasets in $3-$dimensional Cartesian coordinates: photometric galaxies described as lines of sight (green lines), perpendicular filamentary spines (darkblue spines), non-perpendicular filamentary spines (orange spines), and spectroscopic galaxy lines of sight (red lines). All the drawn objects are located in the distance range of $200-600$~Mpc, and on the sphere, they are located at a maximum distance of $8$ degrees from a randomly chosen filamentary spine of the filament catalogue. The photometric galaxy lines of sight are randomly chosen to be $0.25\%$ and the spectroscopic galaxy lines of sight are a randomly chosen to be $0.5\%$ of those in the selected region. } 
         \label{Data_3Dim_Distances}
   \end{figure*}
\subsection{Analysis in $3$-dimensional Euclidean space}\label{3DimDistanceTheory} 

The redshift estimate for photometric galaxies is only roughly known in the observed catalogue~\citep{beck2016photometric}. To estimate distances in $3-$dimensional space between photometric galaxies and the filamentary spines, the location of the photometric galaxy is described by a line of sight. This means that the galaxy may be located at any point along the line following through its position on the sphere. We also represented the galaxy position with a line of sight following through its spherical position for spectroscopic galaxies.
Figure~\ref{Data_3Dim_Distances} draws a representative graph of the spectroscopic galaxy lines of sight, the photometric galaxy lines of sight, and filamentary spines in the selected region.

To obtain a descriptive representation for the distances between filamentary spines and galaxy lines of sight in $3-$dimensional space, we proceeded as follows.
The smallest distance from a filamentary spine to a line of sight of a galaxy was calculated and was noted by $d(Y_{\mathrm{fil}},X_{\mathrm{gal}_\mathrm{l.o.s}})$. 
This distance was calculated between a filamentary spine and a line of sight going through the spherical position of every galaxy in a viewed sample. The set of these distances was calculated for every spine. Then the distribution of the derived distances from all the viewed filamentary spines was smoothed by a constant kernel size and normalised with the count of these computed distances. The kernel size was chosen to be $0.05$~Mpc for all the cases. Different kernel sizes were tested for smoothing ($0.2$~Mpc and $0.01$~Mpc). The results remained independent of the smoothing width. We note this quantity with the form 
\begin{equation}
\tilde{N}[d(Y_{\mathrm{fil}},X_{\mathrm{gal}_\mathrm{l.o.s}}) ].
\end{equation}

For a reference case, random points were generated on the viewed region of a sphere. They are described as $3-$dimensional lines of sight following through these spherical coordinates. The number and spherical region of the generated randomly distributed points are equal to the set of galaxy data in each respective view. Then the smallest distances from a filamentary spine in the corresponding region to a randomly distributed line of sight was calculated. The distance is noted by $d(Y_{\mathrm{fil}},X_{\mathrm{r.p.}_\mathrm{l.o.s}})$. These distances were calculated for every spine. The smoothing and normalization were made in the same fashion as for the previous case. This quantity is presented by the form 
\begin{equation}
\tilde{N}[d(Y_{\mathrm{fil}},X_{\mathrm{r.p.}_\mathrm{l.o.s}})].
\end{equation}
Subsequently, the kernel-smoothed normalised distribution of distances from the filamentary spines to the set of galaxy lines of sight was divided by the kernel-smoothed normalised distribution of distances from the filamentary spines to the set of random lines of sight. The kernel width in both cases is the same.

We denote with $D(Y_{\mathrm{fil}},X_{\mathrm{gal_{l.o.s}}})$ the calculated quotient
\begin{equation}
D(Y_{\mathrm{fil}},X_{\mathrm{gal_{l.o.s}}})=\frac{\tilde{N}[d(Y_{\mathrm{fil}},X_{\mathrm{gal}_\mathrm{l.o.s}})]}{\tilde{N}[d(Y_{\mathrm{fil}},X_{\mathrm{r.p.}_\mathrm{l.o.s}})]}. 
\end{equation}
The quotient of smoothed normalised distributions $D(Y_{\mathrm{fil}},X_{\mathrm{gal_{l.o.s}}})$ can be interpreted subsequently. The statistically significant values $D(Y_{\mathrm{fil}},X_{\mathrm{gal_{l.o.s}}})>1$ mean that the smallest distances from a galaxy line of sight to every filamentary spine are more frequently distributed at this distance window than in case of randomly distributed lines of sight. This indicates a preferred distance for the galaxy lines of sight and filamentary spines in $3-$dimensional space. Values of $D(X_\mathrm{fil},Y_\mathrm{gal_{l.o.s}})$ around $1$ indicate no significant divergence from the distribution of random line-of-sight distances to the filamentary spine set. 

\section{Application of the summary statistics}\label{results}

This section introduces the results obtained by the analysis tools described in Sects.~\ref{Estimation} and~\ref{3DimDistanceTheory}. Firstly, the results were obtained by the bivariate $J-$function for mapped datasets in a region of a sphere, followed with the results obtained by the derived quotient $D$ in $3-$dimensional Euclidean space. 

\subsection{Associations between filamentary spines and photometric galaxies in a region of a sphere}
\label{Bivariateresults}

In this subsection we present the results obtained with the bivariate $J-$function on a sphere $\mathbb{S}^{2}$. For this we calculated the distances between the investigated sets on the sphere $\mathbb{S}^{2}$, which measures the central angle between points on the sphere. 
The calculated distance between two points $u, v $ on $\mathbb{S}^{2}$ with latitudes $\eta_{u}, \eta_{v}$ and longitudes $\lambda_{u}, \lambda_{v}$ is the great-circle arc distance from $u$ to $v$, which is of the form
\begin{eqnarray}
\Delta(u, v)& =& \arccos(\sin(\lambda_{u})\cdot \sin(\lambda_{v})+ 
                    \nonumber\\
& & +\cos(\lambda_{u})\cdot \cos(\lambda_{v}) \cdot \cos(\eta_{v}-\eta_{u})). 
\end{eqnarray}~
We used the mapped filamentary spines in a sphere, whose average distance of the filamentary spine axis is at the distance range of $200-800$~Mpc. The sets of spectroscopic redshift galaxies and photometric redshift galaxies were limited in the same distance range. 
In Sect.~\ref{tomography} we discussed that the orientation of the filamentary spine might affect the clustering effect. To test whether there might be a systematic effect from differently oriented filamentary spines, the bivariate $J-$ function was calculated for all filaments, perpendicular filaments and non-perpendicular filaments. All subsets taken from the datasets are described in Sect.~\ref{tomography}, and their details are listed in Table~\ref{AlphaSets}.

 \begin{figure}
   \centering
   \includegraphics{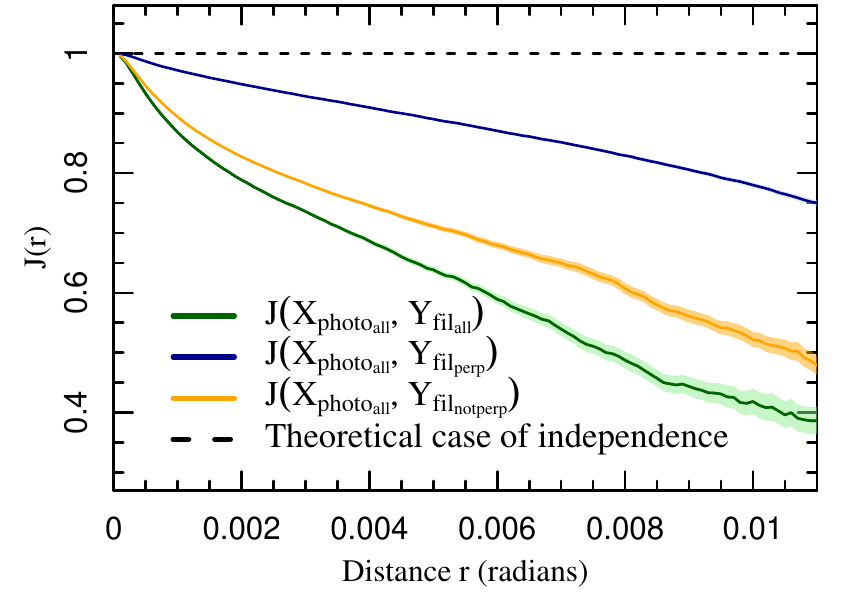} 
     \caption{Results of the bivariate $J-$functions $J(r)_{X_{\mathrm{photo_{all}}}, Y_{\mathrm{fil_{all}}}}$ (green), $J(r)_{X_{\mathrm{photo_{all}}}, X_{\mathrm{fil_{perp}}}}$ (blue), and $J(r)_{X_{\mathrm{photo_{all}}}, X_{\mathrm{fil_{notperp}}}}$ (orange) in comparison with the theoretical reference case representing independence between the studied sets (black dashed line). The radian distance of $0.002$ corresponds to $0.4$~Mpc at the distance of $200$~Mpc and $1.6$~Mpc at the distance of $800$~Mpc.}
         \label{ALLFILS_PERPFILS_NOTPERPFILS}
 \end{figure}

Figure~\ref{ALLFILS_PERPFILS_NOTPERPFILS} draws 
bivariate $J-$functions for photometric galaxies and all filamentary spines in the viewed region and the bivariate $J-$function for the same photometric galaxies and two subsets of the spines, defined as perpendicular to the line of sight and defined as non-perpendicular to the line of sight.
In Fig.~\ref{ALLFILS_PERPFILS_NOTPERPFILS} the decreasing $J(r)_{X_{\mathrm{photo_{all}}}, Y_{\mathrm{fil_{all}}}}$ values below the theoretical reference case represent a positive association between the photometric galaxies and all filamentary spines. The same is shown by the lines of $J(r)_{X_{\mathrm{photo}}, X_{\mathrm{fil_{perp}}}}$ and $J(r)_{X_{\mathrm{photo}}, X_{\mathrm{fil_{notperp}}}}$. This indicates that the photomeric galaxies are positively associated with filaments with different  orientations $\cos(\bar\alpha(Y))$. 
The clustering effect is stronger for all filamentary spines than for perpendicular filaments or non-perpendicular filaments. This is shown by the sharper drop in the $J-$function values for all filamentary spines $J(r)_{X_{\mathrm{photo_{all}}}, Y_{\mathrm{fil_{all}}}}$. 

In Fig.~\ref{ALLFILS_PERPFILS_NOTPERPFILS} the orange line draws the $J-$function for photometric galaxies and filamentary spines that are not perpendicular to the line of sight. The sharper drop in the $J-$function values for every $r$ compared with perpendicular filamentary spines (blue line) denotes a stronger positive association between photometric galaxies and non-perpendicular filamentary spines. The clustering is not as strong as for all filamentary spines, however. All of the drawn bivariate $J-$ functions in Fig.~\ref{ALLFILS_PERPFILS_NOTPERPFILS} indicate a positive association between photometric galaxies and filamentary spines, independent of their orientation towards the line of sight. 

Galaxies tend to reside in groups. Thus the photometric galaxies residing in the groups of the observed spectroscopic galaxies might contribute to a stronger clustering trend between photometric galaxies and filamentary spines because the spine network is detected using a marked point process on the spatial distribution of these spectroscopic galaxies. In the following this problem is addressed in two different ways.

\begin{figure}
   \centering
   \includegraphics{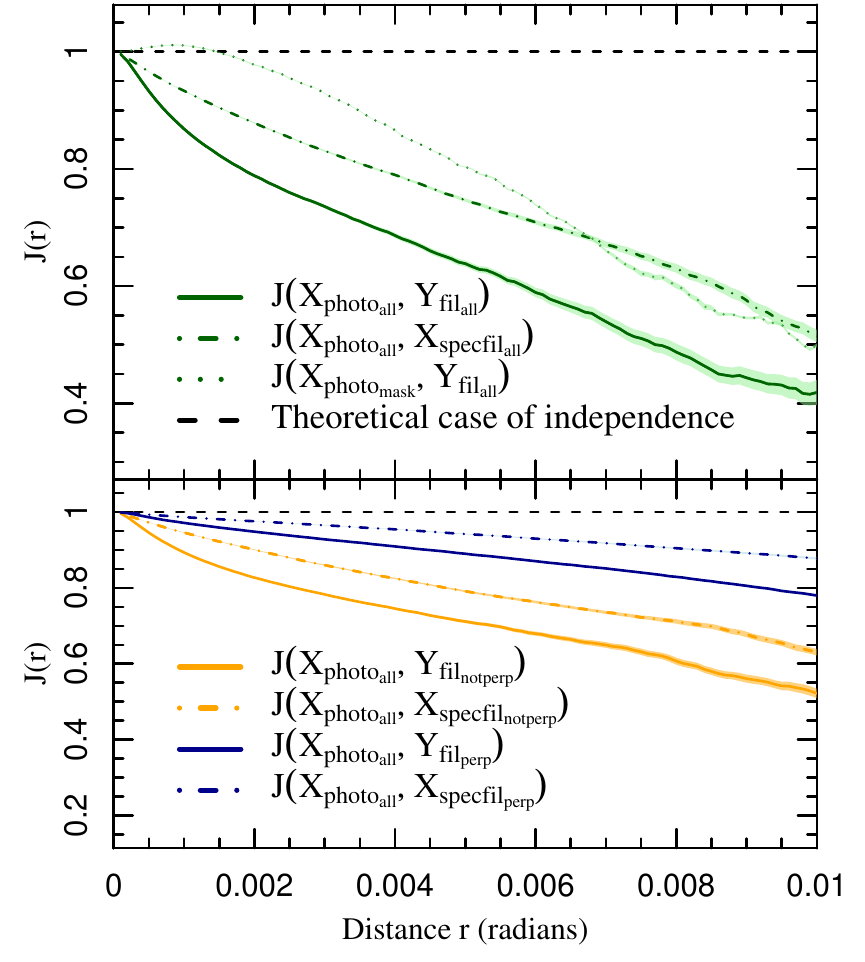}
     \caption{Upper panel: Result of the bivariate $J-$functions $J(r)_{X_{\mathrm{photo_{all}}}, Y_{\mathrm{fil_{all}}}}$ (continuous green line), $J(r)_{X_{\mathrm{photo}}, X_{\mathrm{specfil_{all}}}}$ (dashed green line), and $J(r)_{X_{\mathrm{photo_{mask}}}, Y_{\mathrm{fil_{all}}}}$ (dotted green line) in comparison with the theoretical reference case representing independence between the studied sets (black dashed line). Lower panel: Result of the bivariate $J-$functions $J(r)_{X_{\mathrm{photo_{all}}}, Y_{\mathrm{fil_{perp}}}}$ (blue), $J(r)_{X_{\mathrm{photo}}, X_{\mathrm{specfil_{perp}}}}$ (dashed blue), $J(r)_{X_{\mathrm{photo}}, X_{\mathrm{fil_{notperp}}}}$ (orange),  and $J(r)_{X_{\mathrm{photo}}, X_{\mathrm{specfil_{notperp}}}}$ (dashed orange) in comparison with the theoretical reference case representing independence between the studied sets (black dashed line).}
         \label{SpecsIn_ALLFils_Photoz}
 \end{figure}

We analysed whether the possible clustering between photometric redshift galaxies and filamentary spines shown by the summary statistics might simply come from the clustering effect between photometric redshift galaxies and spectroscopic redshift galaxies inside these filaments. 
We considered a subset of $X_{\mathrm{spec}}$ galaxies, which reside at a certain fixed radius from the filamentary spine they belong to. In the current analysis we fixed it at $0.5$~Mpc, at which a spectroscopic galaxy can be located from its filamentary spine to be named as a galaxy inside the structure. The value of $0.5$~Mpc was chosen based on the method of filamentary spine detection described in~\citet{TempEtAl14} because the radius of the filamentary object was set approximately to the value of $0.5$~Mpc. 
All of the subsets of spectroscopic redshift galaxies inside filaments are within the distance range of $200-800$~Mpc and are located at the maximum distance of $0.5$~Mpc from the filamentary spine axis that the galaxy belongs to. Three sets of these spectroscopic galaxies inside filamentary spines were created: spectroscopic galaxies inside all of the filaments $X_{\mathrm{specfil_{all}}}$, spectroscopic galaxies inside perpendicular filaments $X_{\mathrm{specfil_{perp}}}$ , and spectroscopic galaxies inside non-perpendicular filaments $X_{\mathrm{specfil_{notperp}}}$. The $J-$functions in between the set of photometric galaxies $X_{\mathrm{photo_{all}}}$ and the spectroscopic galaxies subsets described above were calculated and are drawn in Fig.~\ref{SpecsIn_ALLFils_Photoz}. More details about the created subsets of spectroscopic galaxies $X_{\mathrm{specfil}_{\mathrm{all}}}$, $X_{\mathrm{specfil}_{\mathrm{perp}}}$ , and $X_{\mathrm{specfil}_{\mathrm{notperp}}}$ are listed in Table~\ref{AlphaSets}. 

To completely eliminate the effect of the spectroscopic galaxies on the clustering signal between photometric galaxies and filamentary spines, we proceeded as follows. The photometric galaxies were masked out in the observed region of the sphere $\mathbb{S}^{2}$ dependent on the locations of spectroscopic galaxies. This means that we cut ``holes'' in the sphere around the position of every spectroscopic galaxy. The radius of each cut circle was equal to $0.5$~Mpc at the distance of the corresponding spectroscopic galaxy (projected distance on the sky sphere). This resulted in differently sized circular holes in the sphere. For spectroscopic galaxies in the vicinity of the observer (at $200$~Mpc), the $0.5$~Mpc radius region is roughly $0.0025$~radians on the sphere, and for the most distant spectroscopic galaxies ($800$~Mpc) the same radius is roughly $0.000625$~radians on the sphere. Thus photometric galaxies are masked out from the observed region of the sphere dependent on the area around each spectroscopic galaxy in the corresponding view. These masked-out photometric galaxies are noted as $X_{\mathrm{photo}_{\mathrm{mask}}}$. The $J-$function between the $X_{\mathrm{photo}_{\mathrm{mask}}}$ and $Y_{\mathrm{fil_{all}}}$ was calculated and the result is shown in the upper panel of Fig.~\ref{SpecsIn_ALLFils_Photoz}. More details about the subsets of photometric galaxies $X_{\mathrm{photo}_{\mathrm{mask}}}$ we created are listed in Table \ref{AlphaSets}. 

The upper panel of Figure~\ref{SpecsIn_ALLFils_Photoz} shows the bivariate $J-$function for all photometric redshift galaxies $X_{\mathrm{photo_{all}}}$ and all filamentary spines $Y_{\mathrm{fil_{all}}}$ and the bivariate $J-$function for photometric redshift galaxies $X_{\mathrm{photo_{all}}}$ and spectroscopic redshift galaxies inside the same filaments $X_{\mathrm{specfil_{all}}}$. The $J-$function $J(r)_{X_{\mathrm{photo_{all}}}, Y_{\mathrm{fil_{all}}}}$ drops more rapidly and stays below that of $J(r)_{X_{\mathrm{photo}}, X_{\mathrm{specfil_{all}}}}$ for every $r$. This indicates a stronger clustering effect between filamentary spines and photometric galaxies. This result verifies that the clustering effect between photometric galaxies and filaments is truly between the filamentary spines. We also tried different limiting radii ($0.25$ and $1$~Mpc), but  the overall result remained unchanged. 

To remove the possible doubt that the positive association between photometric galaxies and perpendicular filamentary spines might be caused by the photometric redshift galaxies clustering around the spectroscopic redhsift galaxies inside these perpendicular filaments, we drew the $J-$function for ${X_{\mathrm{photo_{perp}}}}$ and $X_{\mathrm{specfil_{perp}}}$. 
In the lower panel of Fig.~\ref{SpecsIn_ALLFils_Photoz} the $J(r)_{X_{\mathrm{photo_{all}}}, Y_{\mathrm{fil_{perp}}}}$ drops more rapidly than $J(r)_{X_{\mathrm{photo}}, X_{\mathrm{specfil_{perp}}}}$ , exhibiting a stronger positive association between the photometric galaxies and perpendicular filamentary spines than in case of photometric galaxies and spectroscopic galaxies inside these filaments. This verifies that the photometric galaxies are positively associated with the filamentary spines and not with the spectroscopic galaxies inside these filaments. The same is shown by comparing $J(r)_{X_{\mathrm{photo_{all}}}, Y_{\mathrm{fil_{notperp}}}}$ with $J(r)_{X_{\mathrm{photo_{all}}}, Y_{\mathrm{specfil_{notperp}}}}$. This confirms that photometric galaxies are more strongly positively associated with non-perpendicular filamentary spines than with the spectroscopic galaxies inhabiting them.

The green dotted line in Fig.~\ref{SpecsIn_ALLFils_Photoz} draws the $J-$function for masked-out photometric galaxies, as was described previously, and all filamentary spines in the corresponding region. The possible cause for the subtle rise in the $J-$function $J(r)_{X_{\mathrm{photo_{mask}}}, Y_{\mathrm{fil_{all}}}}$ values for radians roughly smaller than $0.002$  might be the effect of cutting out $0.5$~Mpc regions around spectroscopic galaxies in their distance to the observer. The photometric galaxy and spectroscopic galaxy clustering radius on the sphere is dependent on the distance of the galaxies. The randomly distributed or arbitrary point locations are independent of the spectroscopic galaxies in the observed region of the sphere and the galaxy distance from the observer. 
The $J-$function shows repulsion between the point process of photometric galaxies and the random set of objects (filamentary spines) for small radian distances. From the great-circle arc distance of $0.002$ radians, the $J-$function denotes a clear clustering effect between the masked-out photometric galaxies and all filamentary spines on the sphere. This indicates that once the radian distance on the sphere exceeds the $0.5$~Mpc distance for all distances of spectroscopic galaxies, the $J-$function values start a steady decline below the value of $1$. Thus after completely eliminating the photometric galaxies that are clustered around spectroscopic galaxies, the rest of the photometric galaxies still show a statistical clustering trend between the filamentary spines and these masked-out photometric galaxies. This indicates that  groups of galaxies, consisting of spectroscopic and photometric galaxies, contribute most of the clustering trend at small distances to the filamentary spines.

\subsection{Studying the distances between filamentary spines and photometric galaxies in $3$-dimensional Euclidean space}\label{3Danalysis}
     
In the previous section \ref{Bivariateresults} we mapped data from $3-$dimensional space to a region of a sphere. The results showed us a positive association between the pairs of projected objects of interest. In this section we estimate the distance distributions between all the described galaxy subsets and the filamentary spine dataset in $3-$dimensional Euclidean space. We calculated the distances from the set of filamentary spine axes to the photometric galaxy lines of sight and the spectroscopic redshift galaxy lines of sights. To do this, a quotient $D$ was built using these distances. The calculation of this quotient is described in depth in the Sect.~\ref{3DimDistanceTheory}. This analysis shows us if the clustering effect can be seen as a physical signal in the smaller distances between the sets. The quotient $D$ for distances from filamentary spines to photometric galaxy lines of sight gives a rough estimate for the radius of the filament according to these galaxies. The detection algorithm described by \citet{TempEtAl14} used a set scale of $\sim 0.5$~Mpc for the radius of the filamentary objects.
   
   \begin{figure}
   \centering
   \includegraphics{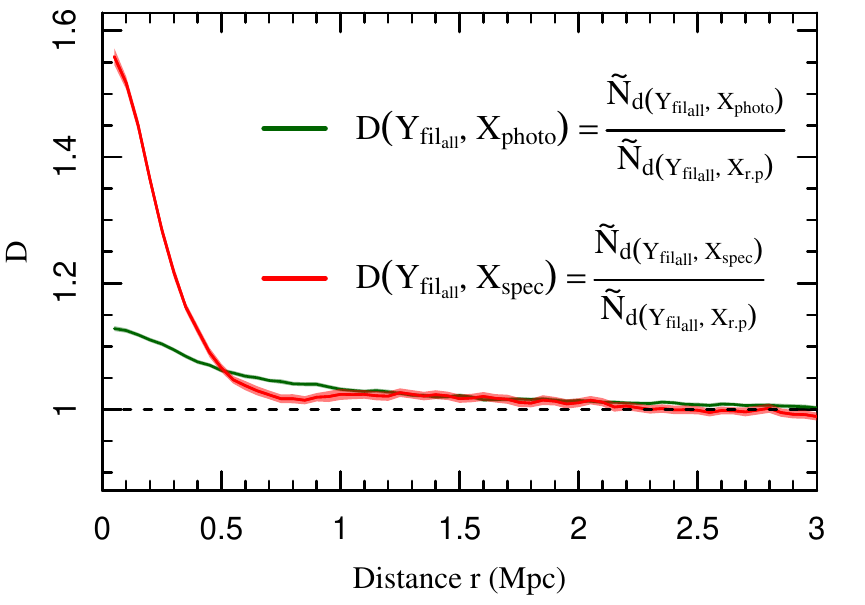}
     \caption{The $y-$axis is the quotient $D$ dependent on the distance $r$, which is described by the $x-$axis. See Sect.~\ref{3DimDistanceTheory} for a detailed description of $\mathrm{D}$. The dark green line draws the quotient of distances from filamentary spines to photometric galaxy lines of sight and to lines of sight that are distributed randomly. The red line draws the quotient of distances from the filamentary spines to spectroscopic galaxy lines of sight and to  lines of sight that are distributed randomly. The shaded areas surrounding the lines denote the calculated $0.95  $ confidence intervals.}
         \label{PHOTOZ_SPECS}
   \end{figure}
  
In Fig. \ref{PHOTOZ_SPECS} the lines depict the quotient, which was calculated for the distances from filamentary spines to galaxy datasets. 
The red line in the upper panel of Fig.~\ref{PHOTOZ_SPECS} shows the strong trend for spectroscopic galaxies. This indicates that the spectroscopic galaxy lines of sight have a significantly greater probability to be located at smaller distances from the filamentary spines than randomly distributed lines of sight are.
This result is highly anticipated as the filament catalogue was derived by applying an object-marked point process with interactions on the same spectroscopic galaxy data.   
The dark green line shows that the photometric galaxy lines of sight are more frequently located at smaller distances from the filamentary spines than the randomly distributed lines of sight. This implies that they carry information about the location of the network. 

Figure~\ref{PHOTOZ_SPECS} also describes the filament sizes. As filaments are described to be filled to a higher degree with galaxies than the surrounding areas, the drop in the relation of the distance distributions illustrates the rough scale of the defined filament. The steep drop around $0.5$~Mpc for the spectroscopic galaxies is again anticipated, as the model was applied on these galaxies with a set radius of about $0.5$~Mpc for the filaments. The decrease of the quotient for the photometric galaxies roughly matches that of spectroscopic galaxies. The photometric galaxies give a rough estimate that the filament widths are about $1$~Mpc. The sharper drop of the quotient (green line) levels somewhat before $1$~Mpc. This confirms that the filament radius is about $0.5$ to $1$~Mpc. The relation of the distance distributions for spectroscopic galaxies (red line) approximately reaches the value $1$ at about the $2.1$~Mpc threshold. The quotient of the distance distributions for photometric galaxies (dark green line) approximately reaches the value $1$ at about the $2.5$~Mpc threshold. 

In this subsection the quotient $D$ was used to estimate the $3$-dimensional distances between the filamentary spines and the two galaxy datasets. The physical distances between photometric galaxies and filamentary spines are probably smaller than $1$~Mpc (Fig.~\ref{PHOTOZ_SPECS}). This shows that the photometric galaxies trace the filamentary network. The value $1$~Mpc is also a rough estimate for the width of a filamentary spine given by the photometric galaxies.

\subsection{Studying the distances between sampled filamentary spines and photometric galaxies in $3$-dimensional Euclidean space} \label{resultstomography}

In Sect.~\ref{tomography} we described all the mechanisms we used to sample the filamentary spine data and the galaxy datasets. For the following analysis we show all the results in the contexts of the sampled objects. 
Firstly, results for the quotient $D$ are shown for filaments that were filtered according to their average angles $\overline{\alpha}$ towards the line of sight.
Then the results for the quotient are drawn, when the objects were sampled into $200$~Mpc wide distance ranges. 
  
Filamentary spines were sorted into sets according to their orientation towards the line of sight $\overline{\alpha}$ (Eq.~\ref{alpha}). More information about the samples is listed in Table~\ref{AlphaSets}. The distances from each of the filamentary spine sample sets to the photometric redshift galaxy lines of sight were calculated.

\begin{figure}
   \centering
   \includegraphics{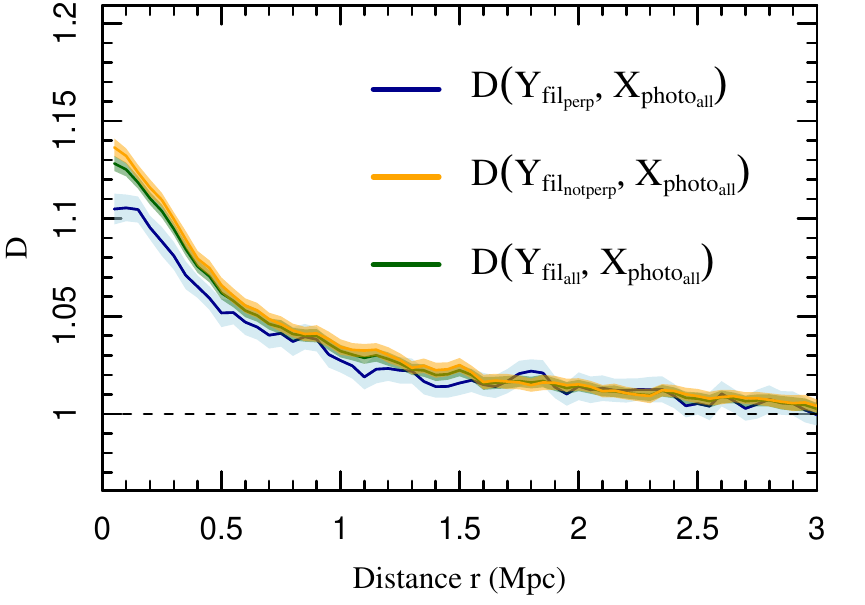}
     \caption{The $y-$axis is the quotient (see Sect.~\ref{3DimDistanceTheory}) value for the subsamples of filamentary spines dependent on the distance $r$, which is described by the $x-$axis. The dark green line draws the distance quotients for all filamentary spines to the photometric galaxy lines of sight. The dark blue line draws the distance quotients for the perpendicular filamentary spines to the photometric galaxy lines of sight. The orange line draws the same quotient for non-perpendicular filamentary spines and photometric galaxy lines of sight. The shaded areas surrounding the lines denote the calculated $0.95  $ confidence intervals.}
         \label{Perp_Notperp}
   \end{figure}

In Fig.~\ref{Perp_Notperp} we show the quotient $D$ of the normalised distance distributions for $\overline{\alpha}$ sampled data. 
Figure~\ref{Perp_Notperp} shows that the photometric galaxy lines of sight are located closer to the two viewed filamentary spine $\overline{\alpha}$ samples more frequently than the randomly distributed lines of sight. This shows the physical trend that distances smaller than $1$~Mpc are more frequent between the two differently oriented filamentary spines and photometric galaxy lines of sights. This verifies that the compression of the Finger-of-God effect with the FoF algorithm does not compress the filamentary spines along the line of sight in such quantities that the trend we studied would be greatly affected. It also indicates that the physical trend for smaller distances between the photometric galaxy lines of sight and filamentary spines is not affected by the orientation of the filamentary spines.

We then selected filamentary spines and photometric galaxies into $200$~Mpc wide distance ranges from $200$ to $800$~Mpc. The filaments were sampled using the mean distance of a filamentary spine to the observer (Eq.~\ref{spinedistance}), and photometric galaxies were sampled using their redshift estimate in the catalogue.
In Sect.~\ref{tomography} the sampling of filamentary spines and photometric galaxies into $200$~Mpc distance ranges is described and justified. The number of objects inside these subsets is given in Table \ref{AlphaSets}.  

\begin{figure}
   \centering
   \includegraphics{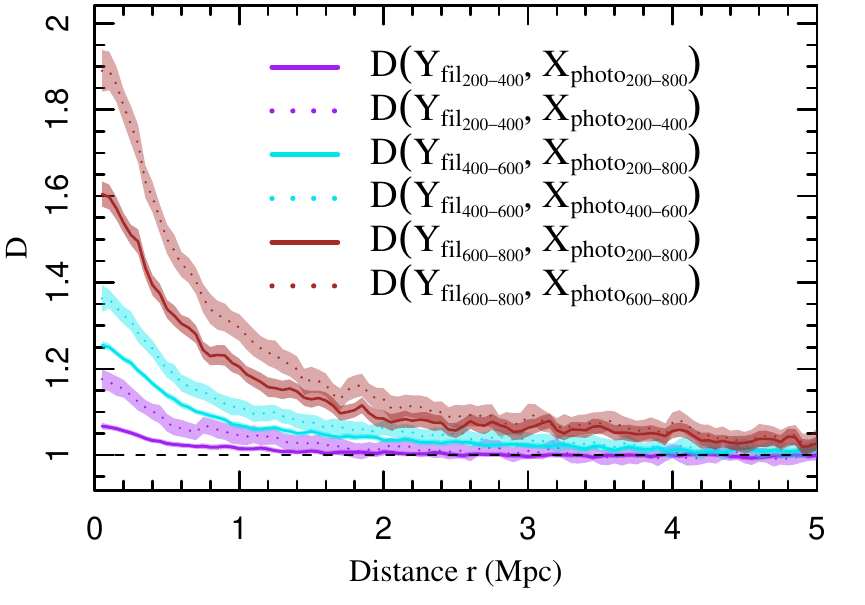}
     \caption{ The $y-$axis is the quotient (see Sect.~\ref{3DimDistanceTheory}) value for the subsamples of data dependent on the distance $r$, which is described by the $x-$axis. The colors purple, turquoise and brown indicate the quotient of distance distributions if the filamentary spine data is filtered into subsequent distance ranges $200-400$~Mpc, $400-600$~Mpc and $600-800$~Mpc. The line type indicates whether the photometric galaxies are filtered into corresponding subsequent ranges or not. For dotted lines both the galaxies and filaments are limited, for continuos lines only filaments are limited to distance ranges and photometric galaxies lie between $200-800$~Mpc. The samples of data are described thoroughly in Table~\ref{AlphaSets}. The shaded areas surrounding the lines denote the calculated $0.95-$confidence intervals.} 
         \label{Tomography}
   \end{figure}

The dotted lines in Fig.~\ref{Tomography} are built of filamentary spines and photometric galaxies sampled into the described $200$~Mpc wide distance ranges. For every calculated quotient, the number of randomly distributed points on the sphere is equal to the number of photometric galaxies in the view. The continuous lines draw the relations, when we did not limit the photometric redshift galaxies into the subsequent distance ranges. In the latter case, the quotients were calculated using photometric galaxies at the distance range of $200-800$~Mpc and filamentary spine sets at the $200-400$~Mpc, $400-600$~Mpc, and $600-800$~Mpc distance ranges. Figure~\ref{Tomography} shows that limiting the photometric redshift galaxies into subsequent ranges enhances the trend of smaller distances between filaments and photometric galaxy lines of sight. In all of the $200$~Mpc wide distance ranges, the trend for smaller distances between the observed sets is stronger when we use photometric redshifts to limit photometric galaxies. This indicates that the photometric redshift carries vital information about the location of the galaxy and might contribute to filamentary structure modelling. 

The quotient $D$ values in Fig.~\ref{Tomography} also show that the trend of smaller distances between filamentary spines and photometric galaxies grows with the distance range. This indicates that the distances between filamentary spines and the lines of sight of photometric galaxies are more frequently smaller for objects in higher distances. The trend of a stronger signal at longer distances holds regardless of whether we apply the limits on photometric galaxies distances. The strongest trend of the quotient indicates more frequent smaller distances between filamentary spines and photometric galaxies lines of sights at the distance range of $600-800$~Mpc. 
One possible explanation for this effect might be that at greater distances, the photometric galaxies we detected are among the brightest and more massive. Galaxies like this are more frequently located in filamentary spines~\citep{2014MNRAS.441.2923C}. 

In this subsection the quotient $D$ was used to analyse the origin of the trend of smaller distances between objects in $3-$dimensions, which we described in the previous subsection \ref{3Danalysis}. The bulk of the trend comes from the objects at greater distances (Fig.~\ref{Tomography}). Limiting photometric galaxies by their redshift estimates and filamentary spines to distance ranges enhances the trend of smaller distances between the objects, which shows that the photometric redshift has viable information about the location of the galaxy.

\section{Conclusion and discussion}\label{conclusion}

We followed a preliminary analysis to obtain insight into the possible contribution of photometric galaxies to the already detected pattern of the filamentary network. The filament catalogue was derived from the spatial distribution of spectroscopic galaxies \citep{TempEtAl14, 2016A&C....16...17T}. This study gave insight into the possible photometric galaxy input to the detection of the filamentary network. The analysis was made in a region of a sphere and in $3-$dimensional Euclidean space. A brief overview of the results is given below. 

In Sect.~\ref{Bivariateresults} the bivariate $J-$functions were computed to analyse possible clustering between photometric galaxies and filamentary spines. The datasets were projected onto a sphere and the summary statistics between the investigated sets was calculated. 
The results of the bivariate $J-$functions in Fig.~\ref{ALLFILS_PERPFILS_NOTPERPFILS} indicate a clear positive association (clustering effect) between photometric galaxies and all the viewed subsamples of filamentary spines. The clustering effect was stronger when the filamentary objects were characterised by their spines rather than by the spectroscopic galaxies that reside in them (Fig.~\ref{SpecsIn_ALLFils_Photoz}). This indicates that the photometric galaxies are truly positively associated with filamentary spines. Photometric galaxies exhibit a tendency to be located in groups with the spectroscopic galaxies. This might have an effect on the clustering trend. This was minimised by masking out all photometric galaxies that were located at a radian distance equivalent to $0.5$~Mpc in the spectroscopic galaxy distance to the observer. The result is also shown in Fig.~\ref{SpecsIn_ALLFils_Photoz}. The effect of cutting ``holes'' into the datasets affected the trend in the range of the sizes of the holes, which reached up to $0.002$ radians in the viewed region of the sphere. The obtained bivariate $J-$function still indicates a clustering effect between the masked-out photometric galaxies and filamentary spines, although for the smallest distances (roughly up to $0.002$ radians), most of the clustering trend is contributed by groups that consist of spectroscopic and photometric redshift galaxies. 

The bivariate $J$-functions in Fig.~\ref{ALLFILS_PERPFILS_NOTPERPFILS} also indicate a change in the strength of the clustering between photometric galaxies and filaments that depends on the filament orientation. Photometric galaxies and filaments noted as non-perpendcicluar to the line of sight exhibit a stronger positive association than photometric galaxies and perpendicular filaments. This can be partly attributed to a roughly $\text{four}$fold difference between the compared filament subset sizes, which is shown in Table~\ref{AlphaSets}.

In Sect.~\ref{3Danalysis} the analysis was made in $3-$dimensional Euclidean space. For this, galaxies in the two datasets were described by lines of sight following through their position on a sphere and the filamentary spines were described as configurations of connected segments in $3-$dimensions. Distances between the filamentary spines and galaxy lines of sight were calculated. For each of the cases, randomly distributed line of sights were generated, and the distances from the viewed filaments to these objects were calculated. Correspondingly, a quotient $D$ was built (Sect.~\ref{3DimDistanceTheory}). This quotient compares the distance distribution of filaments to galaxies to that of filaments to randomly distributed points. The results obtained in this section (Fig.~\ref{PHOTOZ_SPECS})  confirmed the phenomenon that photometric galaxies are located close to the filamentary spines. The distances from filamentary spines to photometric galaxy lines of sight are distributed, when compared to randomly distributed lines of sight, significantly more frequently at smaller distances. This distribution also gives a rough estimate for the filamentary spine width: the radius of the spine is estimated to be roughly around $1$~Mpc. In modelling the filamentary spines from the spatial distribution of spectroscopic galaxies, the filament radius was set to approximately $0.5$~Mpc. 

In Sec.~\ref{resultstomography} the analysis continued between the previously described objects in $3-$dimensions. Filaments and galaxies were additionally sorted either by their angle towards the line of sight and/or distance. The result in Fig.~\ref{Perp_Notperp} showed no clear change in the quotient values, when the filamentary spines were sampled according to their average angle $\bar{\alpha}$ towards the line of sight. Then the spines and photometric galaxies were sampled into $200$~Mpc wide distance ranges. The result in Fig. \ref{Tomography} showed that by limiting photometric galaxies to distance ranges along with the filamentary spines, we were able to boost the trend that they are located close to the structure spines. This shows that the quite uncertain redshift estimates in the catalogue of photometric galaxies already carries vital information about the location of the galaxies.
In addition, the trend of smaller distances between the objects grew with the distance range. This indicates that the further detected photometric galaxies are more likely to be located in filamentary structures. This effect might arise from the detection of more massive galaxies at far away distances. At large distances from the observer, the detected galaxies are more probably in higher density regions of the Universe, such as filamentary spines.

Our analysis verifies that the information hidden in photometric redshift galaxies may contribute to the filamentary network detection. The photometric galaxies give additional information about the structure of the cosmic web. Their highest importance in the detection of large-scale structures is at high redshifts. At lower redshifts, their contribution could be in filling in the ``gaps'' in the currently constructed large-scale map. By complementing the catalogue of spectroscopic redshift galaxies with the photometric redshift galaxies, a more detailed and far-reaching map of the galaxy filaments was constructed. Additionally, in 3-dimensions, the photometric galaxy lines of sight distances from filamentary spines gave a rough estimate that the radius of filaments is about $1$~Mpc. The strongest contribution to the clustering effect seen between photometric galaxies and filamentary spines drawn by the $J-$function at the smallest great-circle arc distances (up to $0.002$ radians) comes from the effect of galaxy groups because the photometric galaxies are clustered around the spectroscopic galaxies. The effect of galaxy groups was also detected in the $3-$dimensional view of the objects: with larger distance, the physical signal of photometric galaxies located close to the spines increased. This was explained by the effect that brighter and more massive galaxies are more probably detected at greater distances. These galaxies are also mostly located in higher density regions of the Universe. This indicates that they might also be galaxies in groups.

Large-scale photometric redshift estimates, although they have been quite imprecise so far, might contribute substantially to the detection or description of the complex filamentary structure of the Universe. The use of these photometric galaxies redshifts would be more complex than the spectroscopic galaxy redshifts because they are measured with lower accuracy. In the upcoming photometric redshift catalogues \citep{benitez2014j}, however, their contribution to the detection of the network would be greater as the redshift estimates are significantly better than in the catalogue used in this analysis (SDSS DR 12), which would make the use of the photometric redshifts more beneficial. Alternatively, the filamentary network could be used to improve photometric redshift estimates, as described in \citet{2015MNRAS.454..463A}. The next step in our study would be to combine the photometric redshift estimates (J-PAS data) and the spectroscopic redshifts (SDSS data) to create a more complete and detailed map of the cosmic web. 

\begin{acknowledgements}

We thank Antti Tamm, Jukka Nevalainen, and Punyakoti Ganeshaiah Veena for valuable comments. We acknowledge the financial support by the institutional research funding IUT26-2 and IUT40-2 of the Estonian Ministry of Education and Research and the support by the Centre of Excellence “Dark side of the Universe” (TK133), which is financed by the European Union through the European Regional Development Fund. ET acknowledges the grant MOBTP86. The work of Radu Stoica was supported by a grant of the Romanian Ministry of National Education and Scientific Research, RDI Program for Space Technology and Advanced Research - STAR, project number $513$.

\end{acknowledgements}

\bibliographystyle{aa} 
\bibliography{aa35096-19}

\begin{thebibliography}{48}
\expandafter\ifx\csname natexlab\endcsname\relax\def\natexlab#1{#1}\fi

\bibitem[{{Aragon-Calvo} {et~al.}(2015){Aragon-Calvo}, {van de Weygaert},
  {Jones}, \& {Mobasher}}]{2015MNRAS.454..463A}
{Aragon-Calvo}, M.~A., {van de Weygaert}, R., {Jones}, B. J.~T., \& {Mobasher},
  B. 2015, \mnras, 454, 463

\bibitem[{{Baddeley} {et~al.}(2007){Baddeley}, {B{\'a}r{\'a}ny}, \&
  {Schneider}}]{baddeley2007spatial}
{Baddeley}, A., {B{\'a}r{\'a}ny}, I., \& {Schneider}, R. 2007, Stochastic
  Geometry: Lectures given at the CIME Summer School held in Martina Franca,
  Italy, September 13--18, 2004, 1

\bibitem[{{Baddeley} {et~al.}(2015){Baddeley}, {Rubak}, \&
  {Turner}}]{baddeley2015spatial}
{Baddeley}, A., {Rubak}, E., \& {Turner}, R. 2015, Spatial point patterns:
  methodology and applications with R (CRC Press)

\bibitem[{{Beck} {et~al.}(2016){Beck}, {Dobos}, {Budav{\'a}ri}, {Szalay}, \&
  {Csabai}}]{beck2016photometric}
{Beck}, R., {Dobos}, L., {Budav{\'a}ri}, T., {Szalay}, A.~S., \& {Csabai}, I.
  2016, \mnras, 460, 1371

\bibitem[{{Benitez} {et~al.}(2014){Benitez}, {Dupke}, {Moles}, {Sodre},
  {Cenarro}, {Marin-Franch}, {Taylor}, {Cristobal}, {Fernandez-Soto}, {Mendes
  de Oliveira}, {Cepa-Nogue}, {Abramo}, {Alcaniz}, {Overzier},
  {Hernandez-Monteagudo}, {Alfaro}, {Kanaan}, {Carvano}, {Reis}, {Martinez
  Gonzalez}, {Ascaso}, {Ballesteros}, {Xavier}, {Varela}, {Ederoclite},
  {Vazquez Ramio}, {Broadhurst}, {Cypriano}, {Angulo}, {Diego}, {Zandivarez},
  {Diaz}, {Melchior}, {Umetsu}, {Spinelli}, {Zitrin}, {Coe}, {Yepes}, {Vielva},
  {Sahni}, {Marcos- Caballero}, {Shu Kitaura}, {Maroto}, {Masip}, {Tsujikawa},
  {Carneiro}, {Gonzalez Nuevo}, {Carvalho}, {Reboucas}, {Carvalho}, {Abdalla},
  {Bernui}, {Pigozzo}, {Ferreira}, {Chandrachani Devi}, {Bengaly}, {Campista},
  {Amorim}, {Asari}, {Bongiovanni}, {Bonoli}, {Bruzual}, {Cardiel}, {Cava},
  {Cid Fernandes}, {Coelho}, {Cortesi}, {Delgado}, {Diaz Garcia}, {Espinosa},
  {Galliano}, {Gonzalez-Serrano}, {Falcon- Barroso}, {Fritz}, {Fernandes},
  {Gorgas}, {Hoyos}, {Jimenez-Teja}, {Lopez-Aguerri}, {Lopez-San Juan},
  {Mateus}, {Molino}, {Novais}, {OMill}, {Oteo}, {Perez- Gonzalez},
  {Poggianti}, {Proctor}, {Ricciardelli}, {Sanchez-Blazquez}, {Storchi-
  Bergmann}, {Telles}, {Schoennell}, {Trujillo}, {Vazdekis}, {Viironen},
  {Daflon}, {Aparicio-Villegas}, {Rocha}, {Ribeiro}, {Borges}, {Martins},
  {Marcolino}, {Martinez-Delgado}, {Perez-Torres}, {Siffert}, {Calvao}, {Sako},
  {Kessler}, {Alvarez-Candal}, {De Pra}, {Roig}, {Lazzaro}, {Gorosabel}, {Lopes
  de Oliveira}, {Lima-Neto}, {Irwin}, {Liu}, {Alvarez}, {Balmes}, {Chueca},
  {Costa-Duarte}, {da Costa}, {Dantas}, {Diaz}, {Fabregat}, {Ferrari},
  {Gavela}, {Gracia}, {Gruel}, {Gutierrez}, {Guzman}, {Hernandez- Fernandez},
  {Herranz}, {Hurtado-Gil}, {Jablonsky}, {Laporte}, {Le Tiran}, {Licandro},
  {Lima}, {Martin}, {Martinez}, {Montero}, {Penteado}, {Pereira}, {Peris},
  {Quilis}, {Sanchez-Portal}, {Soja}, {Solano}, {Torra}, \&
  {Valdivielso}}]{benitez2014j}
{Benitez}, N., {Dupke}, R., {Moles}, M., {et~al.} 2014, arXiv e-prints,
  arXiv:1403.5237

\bibitem[{{Beygu} {et~al.}(2017){Beygu}, {Peletier}, {van der Hulst},
  {Jarrett}, {Kreckel}, {van de Weygaert}, {van Gorkom}, \&
  {Aragon-Calvo}}]{2017MNRAS.464..666B}
{Beygu}, B., {Peletier}, R.~F., {van der Hulst}, J.~M., {et~al.} 2017, \mnras,
  464, 666

\bibitem[{{Bond} {et~al.}(1996){Bond}, {Kofman}, \&
  {Pogosyan}}]{1996Natur.380..603B}
{Bond}, J.~R., {Kofman}, L., \& {Pogosyan}, D. 1996, \nat, 380, 603

\bibitem[{{Cautun} {et~al.}(2014){Cautun}, {van de Weygaert}, {Jones}, \&
  {Frenk}}]{2014MNRAS.441.2923C}
{Cautun}, M., {van de Weygaert}, R., {Jones}, B. J.~T., \& {Frenk}, C.~S. 2014,
  \mnras, 441, 2923

\bibitem[{{Chiu} {et~al.}(2013){Chiu}, {Stoyan}, {Kendall}, \&
  {Mecke}}]{ChiuEtAl13}
{Chiu}, S.~N., {Stoyan}, D., {Kendall}, W.~S., \& {Mecke}, J. 2013, Stochastic
  Geometry and its Applications. Third Edition. (John Wiley and Sons)

\bibitem[{{Colless} {et~al.}(2001){Colless}, {Dalton}, {Maddox}, {Sutherland },
  {Norberg}, {Cole}, {Bland -Hawthorn}, {Bridges}, {Cannon}, {Collins},
  {Couch}, {Cross}, {Deeley}, {De Propris}, {Driver}, {Efstathiou}, {Ellis},
  {Frenk}, {Glazebrook}, {Jackson}, {Lahav}, {Lewis}, {Lumsden}, {Madgwick},
  {Peacock}, {Peterson}, {Price}, {Seaborne}, \&
  {Taylor}}]{2001MNRAS.328.1039C}
{Colless}, M., {Dalton}, G., {Maddox}, S., {et~al.} 2001, \mnras, 328, 1039

\bibitem[{{Crone Odekon} {et~al.}(2018){Crone Odekon}, {Hallenbeck}, {Haynes},
  {Koopmann}, {Phi}, \& {Wolfe}}]{odekon2017effect}
{Crone Odekon}, M., {Hallenbeck}, G., {Haynes}, M.~P., {et~al.} 2018, \apj,
  852, 142

\bibitem[{{de Lapparent} {et~al.}(1986){de Lapparent}, {Geller}, \&
  {Huchra}}]{1986ApJ...302L...1D}
{de Lapparent}, V., {Geller}, M.~J., \& {Huchra}, J.~P. 1986, \apj, 302, L1

\bibitem[{{Foxall} \& {Baddeley}(2002)}]{foxall2002nonparametric}
{Foxall}, R. \& {Baddeley}, A. 2002, Journal of the Royal Statistical Society:
  Series C (Applied Statistics), 51, 165

\bibitem[{{Ganeshaiah Veena} {et~al.}(2018){Ganeshaiah Veena}, {Cautun}, {van
  de Weygaert}, {Tempel}, {Jones}, {Rieder}, \& {Frenk}}]{2018MNRAS.481..414G}
{Ganeshaiah Veena}, P., {Cautun}, M., {van de Weygaert}, R., {et~al.} 2018,
  \mnras, 481, 414

\bibitem[{{Gott} {et~al.}(1986){Gott}, {Melott}, \&
  {Dickinson}}]{1986ApJ...306..341G}
{Gott}, J.~Richard, I., {Melott}, A.~L., \& {Dickinson}, M. 1986, \apj, 306,
  341

\bibitem[{{Gott} {et~al.}(2009){Gott}, {Choi}, {Park}, \&
  {Kim}}]{2009ApJ...695L..45G}
{Gott}, J.~R., {Choi}, Y.-Y., {Park}, C., \& {Kim}, J. 2009, \apj, 695, L45

\bibitem[{{Huchra} {et~al.}(1983){Huchra}, {Davis}, \&
  {Latham}}]{1983sgrd.book.....H}
{Huchra}, J., {Davis}, M., \& {Latham}, D. 1983, {A survey of galaxy redshifts.
  IV: The data}

\bibitem[{{Illian} {et~al.}(2008){Illian}, {Penttinen}, {Stoyan}, \&
  {Stoyan}}]{illian2008statistical}
{Illian}, J., {Penttinen}, A., {Stoyan}, H., \& {Stoyan}, D. 2008, Statistical
  analysis and modelling of spatial point patterns, Vol.~70 (John Wiley \&
  Sons)

\bibitem[{{J{\~o}eveer} {et~al.}(1978){J{\~o}eveer}, {Einasto}, \&
  {Tago}}]{joeveer1978spatial}
{J{\~o}eveer}, M., {Einasto}, J., \& {Tago}, E. 1978, \mnras, 185, 357

\bibitem[{{Jackson}(1972)}]{1972MNRAS.156P...1J}
{Jackson}, J.~C. 1972, \mnras, 156, 1P

\bibitem[{{Jones}(2017)}]{2017prcobookBJ}
{Jones}, B.~J.~T. 2017, {Precision Cosmology}

\bibitem[{{Kleinschroth} {et~al.}(2016){Kleinschroth}, {Healey},
  {Gourlet-Fleury}, {Mortier}, \& {Stoica}}]{kleinschroth2016effects}
{Kleinschroth}, F., {Healey}, J.~R., {Gourlet-Fleury}, S., {Mortier}, F., \&
  {Stoica}, R.~S. 2016, arXiv e-prints, arXiv:1602.01436

\bibitem[{{Kuutma} {et~al.}(2017){Kuutma}, {Tamm}, \& {Tempel}}]{Kuutma2017}
{Kuutma}, T., {Tamm}, A., \& {Tempel}, E. 2017, \aap, 600, L6

\bibitem[{{Lawrence} {et~al.}(1999){Lawrence}, {Rowan-Robinson}, {Ellis},
  {Frenk}, {Efstathiou}, {Kaiser}, {Saunders}, {Parry}, {Xiaoyang}, \&
  {Crawford}}]{1999MNRAS.308..897L}
{Lawrence}, A., {Rowan-Robinson}, M., {Ellis}, R.~S., {et~al.} 1999, \mnras,
  308, 897

\bibitem[{{Libeskind} {et~al.}(2018){Libeskind}, {van de Weygaert}, {Cautun},
  {Falck}, {Tempel}, {Abel}, {Alpaslan}, {Arag{\'o}n-Calvo}, {Forero-Romero},
  {Gonzalez}, {Gottl{\"o}ber}, {Hahn}, {Hellwing}, {Hoffman}, {Jones},
  {Kitaura}, {Knebe}, {Manti}, {Neyrinck}, {Nuza}, {Padilla}, {Platen},
  {Ramachandra}, {Robotham}, {Saar}, {Shandarin}, {Steinmetz}, {Stoica},
  {Sousbie}, \& {Yepes}}]{noam2017tempel}
{Libeskind}, N.~I., {van de Weygaert}, R., {Cautun}, M., {et~al.} 2018, \mnras,
  473, 1195

\bibitem[{{Martinez} \& {Saar}(2001)}]{martinez2001statistics}
{Martinez}, V.~J. \& {Saar}, E. 2001, Statistics of the galaxy distribution
  (CRC Press)

\bibitem[{{M{\o}ller} \& {Waagepetersen}(2004)}]{MollWaag04}
{M{\o}ller}, J. \& {Waagepetersen}, R.~P. 2004, Statistical inference and
  simulation for spatial point processes (Chapman and Hall/CRC, Boca Raton)

\bibitem[{{Moore} {et~al.}(1992){Moore}, {Frenk}, {Weinberg}, {Saunders},
  {Lawrence}, {Ellis}, {Kaiser}, {Efstathiou}, \&
  {Rowan-Robinson}}]{1992MNRAS.256..477M}
{Moore}, B., {Frenk}, C.~S., {Weinberg}, D.~H., {et~al.} 1992, \mnras, 256, 477

\bibitem[{{Myllym{\"{a}}ki} {et~al.}(2017){Myllym{\"{a}}ki}, {Mrkvi{\v{c}}ka},
  {Grabarnik}, {Seijo}, \& {Hahn}}]{MyllEtAl17}
{Myllym{\"{a}}ki}, M., {Mrkvi{\v{c}}ka}, T., {Grabarnik}, P., {Seijo}, H., \&
  {Hahn}, U. 2017, Journal of the Royal Statistical Society, Series B

\bibitem[{{Peebles}(1980)}]{peebles1980large}
{Peebles}, P.~J.~E. 1980, {The large-scale structure of the universe}

\bibitem[{{Pimbblet} {et~al.}(2005){Pimbblet}, {Edge}, \&
  {Couch}}]{2005MNRAS.357L..45P}
{Pimbblet}, K.~A., {Edge}, A.~C., \& {Couch}, W.~J. 2005, \mnras, 357, L45

\bibitem[{{Stoica}(2010)}]{StoicaMarkedPoints}
{Stoica}, R.~S. 2010, European Physical Journal Special Topics, 186, 123

\bibitem[{{Stoica} {et~al.}(2010){Stoica}, {Mart{\'\i}nez}, \&
  {Saar}}]{Stoica2010}
{Stoica}, R.~S., {Mart{\'\i}nez}, V.~J., \& {Saar}, E. 2010, \aap, 510, A38

\bibitem[{{Sunyaev} \& {Zeldovich}(1970)}]{Sunyaev1969}
{Sunyaev}, R.~A. \& {Zeldovich}, Y.~B. 1970, \apss, 7, 3

\bibitem[{{Tempel} {et~al.}(2011){Tempel}, {Saar}, {Liivam{\"a}gi}, {Tamm},
  {Einasto}, {Einasto}, \& {M{\"u}ller}}]{2011A&A...529A..53T}
{Tempel}, E., {Saar}, E., {Liivam{\"a}gi}, L.~J., {et~al.} 2011, \aap, 529, A53

\bibitem[{{Tempel} {et~al.}(2016){Tempel}, {Stoica}, {Kipper}, \&
  {Saar}}]{2016A&C....16...17T}
{Tempel}, E., {Stoica}, R.~S., {Kipper}, R., \& {Saar}, E. 2016, Astronomy and
  Computing, 16, 17

\bibitem[{{Tempel} {et~al.}(2014{\natexlab{a}}){Tempel}, {Stoica},
  {Mart{\'\i}nez}, {Liivam{\"a}gi}, {Castellan}, \& {Saar}}]{TempEtAl14}
{Tempel}, E., {Stoica}, R.~S., {Mart{\'\i}nez}, V.~J., {et~al.}
  2014{\natexlab{a}}, \mnras, 438, 3465

\bibitem[{{Tempel} {et~al.}(2012){Tempel}, {Tago}, \&
  {Liivam{\"a}gi}}]{TempelGroups2012}
{Tempel}, E., {Tago}, E., \& {Liivam{\"a}gi}, L.~J. 2012, \aap, 540, A106

\bibitem[{{Tempel} {et~al.}(2014{\natexlab{b}}){Tempel}, {Tamm}, {Gramann},
  {Tuvikene}, {Liivam{\"a}gi}, {Suhhonenko}, {Kipper}, {Einasto}, \&
  {Saar}}]{2014A&A...566A...1T}
{Tempel}, E., {Tamm}, A., {Gramann}, M., {et~al.} 2014{\natexlab{b}}, \aap,
  566, A1

\bibitem[{{Tempel} {et~al.}(2017){Tempel}, {Tuvikene}, {Kipper}, \&
  {Libeskind}}]{2017A&A...602A.100T}
{Tempel}, E., {Tuvikene}, T., {Kipper}, R., \& {Libeskind}, N.~I. 2017, \aap,
  602, A100

\bibitem[{{Tully} \& {Fisher}(1978)}]{FingerOfGodTully}
{Tully}, R.~B. \& {Fisher}, J.~R. 1978, in IAU Symposium, Vol.~79, Large Scale
  Structures in the Universe, ed. M.~S. {Longair} \& J.~{Einasto}, 31

\bibitem[{{Turner} \& {Gott}(1976)}]{TurnerFOF}
{Turner}, E.~L. \& {Gott}, J.~R., I. 1976, The Astrophysical Journal Supplement
  Series, 32, 409

\bibitem[{{van Lieshout}(2000)}]{Lies00}
{van Lieshout}, M.~N.~M. 2000, Markov Point Processes and their Applications
  (Imperial College Press, London)

\bibitem[{{van Lieshout}(2016)}]{2016arXiv160402302V}
{van Lieshout}, M.~N.~M. 2016, arXiv e-prints, arXiv:1604.02302

\bibitem[{{van Lieshout} \& {Baddeley}(1996)}]{lieshout1996nonparametric}
{van Lieshout}, M.~N.~M. \& {Baddeley}, A.~J. 1996, Statistica Neerlandica, 50,
  344

\bibitem[{{van Lieshout} \& {Baddeley}(1999)}]{LiesBadd99}
{van Lieshout}, M.~N.~M. \& {Baddeley}, A.~J. 1999, Scandinavian Journal of
  Statistics, 26, 511

\bibitem[{{Vogeley} {et~al.}(1994){Vogeley}, {Park}, {Geller}, {Huchra}, \&
  {Gott}}]{1994ApJ...420..525V}
{Vogeley}, M.~S., {Park}, C., {Geller}, M.~J., {Huchra}, J.~P., \& {Gott},
  J.~Richard, I. 1994, \apj, 420, 525

\bibitem[{{York} {et~al.}(2000){York}, {Adelman}, {Anderson}, {Anderson},
  {Annis}, {Bahcall}, {Bakken}, {Barkhouser}, {Bastian}, {Berman}, {Boroski},
  {Bracker}, {Briegel}, {Briggs}, {Brinkmann}, {Brunner}, {Burles}, {Carey},
  {Carr}, {Castander}, {Chen}, {Colestock}, {Connolly}, {Crocker}, {Csabai},
  {Czarapata}, {Davis}, {Doi}, {Dombeck}, {Eisenstein}, {Ellman}, {Elms},
  {Evans}, {Fan}, {Federwitz}, {Fiscelli}, {Friedman}, {Frieman}, {Fukugita},
  {Gillespie}, {Gunn}, {Gurbani}, {de Haas}, {Haldeman}, {Harris}, {Hayes},
  {Heckman}, {Hennessy}, {Hindsley}, {Holm}, {Holmgren}, {Huang}, {Hull},
  {Husby}, {Ichikawa}, {Ichikawa}, {Ivezi{\'c}}, {Kent}, {Kim}, {Kinney},
  {Klaene}, {Kleinman}, {Kleinman}, {Knapp}, {Korienek}, {Kron}, {Kunszt},
  {Lamb}, {Lee}, {Leger}, {Limmongkol}, {Lindenmeyer}, {Long}, {Loomis},
  {Loveday}, {Lucinio}, {Lupton}, {MacKinnon}, {Mannery}, {Mantsch}, {Margon},
  {McGehee}, {McKay}, {Meiksin}, {Merelli}, {Monet}, {Munn}, {Narayanan},
  {Nash}, {Neilsen}, {Neswold}, {Newberg}, {Nichol}, {Nicinski}, {Nonino},
  {Okada}, {Okamura}, {Ostriker}, {Owen}, {Pauls}, {Peoples}, {Peterson},
  {Petravick}, {Pier}, {Pope}, {Pordes}, {Prosapio}, {Rechenmacher}, {Quinn},
  {Richards}, {Richmond}, {Rivetta}, {Rockosi}, {Ruthmansdorfer}, {Sandford},
  {Schlegel}, {Schneider}, {Sekiguchi}, {Sergey}, {Shimasaku}, {Siegmund},
  {Smee}, {Smith}, {Snedden}, {Stone}, {Stoughton}, {Strauss}, {Stubbs},
  {SubbaRao}, {Szalay}, {Szapudi}, {Szokoly}, {Thakar}, {Tremonti}, {Tucker},
  {Uomoto}, {Vanden Berk}, {Vogeley}, {Waddell}, {Wang}, {Watanabe},
  {Weinberg}, {Yanny}, {Yasuda}, \& {SDSS Collaboration}}]{2000SDSS}
{York}, D.~G., {Adelman}, J., {Anderson}, John~E., J., {et~al.} 2000, \aj, 120,
  1579

\end{thebibliography}

\end{document}